\documentclass[12pt]{article}
\usepackage{latexsym}
\usepackage{amssymb}
\usepackage{amsmath}
\usepackage{graphicx}
\usepackage{bbm}
\usepackage{enumitem}
\usepackage{slashed}
\usepackage{hyperref}
\usepackage{xcolor}
\usepackage[titletoc,title]{appendix}

\def\be{\begin{equation}}
\def\ee{\end{equation}}
\newcommand\bra[1]{{\langle {#1}|}}
\newcommand\ket[1]{{|{#1}\rangle}}
\def\Tr{{\rm Tr}}
\def\dd{\mbox{d}}

\def\om{\omega}
\def\bra{\langle}
\def\ket{\rangle}
\def\a{\alpha}

\def\G{\Gamma}

\def\f{\phi}

\def\k{\kappa}

\def\L{\Lambda}

\def\t{\tau}

\def\pa{\partial}

\newcommand{\sm}[1]{\mbox{\scriptsize #1}}
\newcommand{\tn}[1]{\mbox{\tiny #1}}
\renewcommand{\@}[1]{\sqrt{#1}}
\renewcommand{\le}[1]{\label{#1}\end{eqnarray}}
\newcommand{\bea}{\begin{eqnarray}}
\newcommand{\eea}{\end{eqnarray}}
\newcommand{\eq}[1]{(\ref{#1})}
\def\nn{\nonumber\\}

\def\half{{1\over2}\,}

\begin{document}
\pagestyle{plain}

\centerline{{\Large \bf The Geometric View of Theories}}
\vskip.5cm
\centerline{{\Large\bf }}
\vskip.7cm

\begin{center}
{\large Sebastian De Haro}\\
\vskip .7truecm
{\it Institute for Logic, Language and Computation, University of Amsterdam}\\
{\it Institute of Physics, University of Amsterdam}\\

\end{center}

\vskip .7truecm

\begin{center}
\today
\end{center}

\vskip 4truecm

\begin{center}
\textbf{\large \bf Abstract}
\end{center}

Recent critiques of the semantic conception of scientific theories suggest that a theory is not best formulated as a collection of models satisfying some set of kinematical or dynamical conditions. Thus it has been argued that additional structure on the set of models is required. Furthermore, there are calls for developing a `theory of theories', where what was formerly a `theory' is seen as a `model' within a larger theoretical structure. This paper makes a two-pronged proposal for the ``shape'' that physical theories should take, based on recent insights on dualities and quasi-dualities in physics. First, I develop a geometric view of theories, according to which a physical theory is a set of models equipped with topological and geometric structure. This general view is briefly illustrated in an example from quantum cosmology. Second, I make a more specific proposal for a natural structure that can encompass various `theories' as its models, with topological and algebraic-geometric structure on them. I call the latter more specific structure a `model bundle', where the models are in the fibres and there is a moduli space in the base. I illustrate my second proposal in an example from quantum field theory. This view highlights the important role of quasi-dualities as local transition functions between fibres; dualities are recovered as global transition functions when the bundle is trivial. I discuss some philosophical issues that this geometric view of physical theories opens up, such as its realist interpretation.

\vskip1.5cm
\noindent Published in {\it Synthese} (2026), 207, 37, pp.~1--39.

\newpage

\tableofcontents

\newpage

\section{Introduction}\label{intro}

One of the central questions in the philosophy of science has traditionally concerned the structure of scientific theories: Is a scientific theory best described as a collection of sentences in a formal language; or, rather, as a set of models, each of them defined as a set-theoretic structure? These two views are, of course, respectively, the syntactic and semantic conceptions of theories.

Recent discussions have emphasised the kinship between these two views: models were always integral to the syntactic view, and likewise the semantic view was---despite some radical-sounding initial claims by its proponents---never meant to dispense with language. In fact, one speaks of a recent consensus, according to which both aspects are needed. I will call this consensus view the `combined view':\footnote{See Halvorson (2012), Glymour (2013), van Fraassen (2014:~p.~276), and Lutz (2017:~p.~330). An overview of the subject is given in Frigg (2023), who discusses the consensus at p.~167. However, there is also dissent: Wallace's (2022:~p.~349; 2024:~pp.~1--2) distinction between `language-first' and `math-first' approaches to physical theories widens the gap between the syntactic and semantic conceptions. This distinction seems to be chiefly motivated by the aim to develop the formal underpinnings of a structural realist metaphysics.} 
namely, the view that a scientific theory requires both a language and a set of models, where the language need not be first-order logic. On this view, the linguistic aspect includes a commitment to a set of sentences, and each model makes these sentences true.

The recent literature has stressed that even this combined view is insufficient. The evidence for this comes mostly from studying inter-theoretic relations. There are two related points, (i) and (ii), where (ii) is a consequence of (i):\\
\\
(i)~~{\it Inter-theoretic relations:} These suggest that there are formal and structural relationships between models that the standard conceptions of theories usually treat as unrelated. Thus discussions of equivalence of theories emphasise different ways in which models or theories can be equivalent or inequivalent.\footnote{For an overview of these developments, see Weatherall (2019). } 
Likewise, relations of reduction and correspondence suggest the need to develop formal notions of `closeness' and `membership' of theories, so as to have criteria for when a theory approximates, and when it subsumes, another theory. Therefore, to go beyond the combined view, one needs to choose the inter-theoretic relations that one wishes to include as part of one's conception of `theory'. Consequently, these inter-theoretic relations, especially equivalence relations, may prompt a change in one's usage of `theory' and `model'.\\
\\
(ii)~~{\it Structures:} Inter-theoretic relations suggest that there is {\it further structure} on the set of theories and models, i.e.~structure above and beyond the structure that is used to define a model. Because of the need to consider this further structure, the semantic slogan that `a theory is a collection of models' is regarded as insufficient. I shall follow Halvorson (2019:~p.~277) in calling the view that includes this further structure, {\it the structured view of theories}.\footnote{Halvorson (2019:~p.~277--278) dubs the unstructured view of theories a `flat' view. He emphasises that the traditional syntactic and semantic conceptions are both flat, and require structure.}\\

Recent proponents of the structured view of theories include Halvorson (2012:~pp.~204--205; 2019:~p.~277) and Halvorson and Tsementzis (2017:~pp.~413--414).\footnote{See also Curiel (2014:~p.~275; cf.~p.~318) and North (2021:~pp.~187--188). In general relativity, spaces of solutions of the Einstein field equations, and criteria for closeness, were discussed very early on (see Hawking and Ellis, 1973:~pp.~249, 252--254). For a philosophical discussion, see Belot (2018:~p.~968); also Fletcher (2016:~p.~366). For example, Geroch (1969:~p.~182) constructed a five-manifold from a one-parameter family of four-manifolds that depend smoothly on a continuous parameter. His motivation was to define the limit of this sequence of four-dimensional metrics, as the metric on the boundary of the five-dimensional manifold.} They motivate their requirement for having `relevant topological information' on the set of models by the question of when a theory, understood as a set of sentences, can be reconstructed from its category of models. Lehmkuhl (2017:~pp.~1--10) is a programmatic call for a `theory of theories'. The first (modest) step in this programme is to compare individual theories.

Although the combined view officially recognises that the set of models inherits structure from the way the models satisfy the sentences of the theory (and from the logical relations between those sentences), it is seldom made concrete what this structure looks like for actual theories---let alone for physical theories. As we will see from the specifications of points (i) and (ii) suggested by recent work in physics below, the literature on theoretical equivalence only in part addresses point (i), and point (ii) not at all. As I will explain below, these qualifications, `in part' and `not at all', are meant to signal that previous work leaves open the following two important questions: What is the physical motivation for introducing structure on the set of models, and what structure should be considered on a set of models of a physical theory? What is the physical interpretation of this structure?

Lehmkuhl (2017:~pp.~1, 10) has stressed how daunting is the task of finding a `theory of theories', and recommends a `step by step' approach. For, if these questions are to be relevant for actual scientific theories, they cannot be answered by lofty philosophical or logical considerations---useful as these might be for some purposes. As my examples will illustrate, philosophy here needs to take current theoretical and mathematical physics as its allies: it needs to go beyond an engagement limited to ``good old friends'', like classical mechanics and general relativity, and incorporate insights from quantum field theory and quantum gravity, including string theory. And it needs to include other theories like statistical mechanics that have usually not played a central role in discussions of the identity and structure of scientific theories, except for particular questions like reduction.

I will argue that there are two key notions from these areas of physics that need to be incorporated into the philosophical discussion of the structure of physical theories: namely, dualities i.e.~isomorphisms of theories, and their generalizations which I call {\it quasi-dualities} (where the map falls short of being an isomorphism), together with the idea of a {\it moduli space} and, more generally, a space of parameters. Although spaces of theories and moduli spaces may be familiar from quantum gravity and other areas of physics and mathematics: exceptions aside, virtually no trace of these notions can be found in the philosophical literature.\footnote{The only in-depth philosophical discussions of the moduli space of a physical theory that I am aware of are De Haro and Butterfield (2025), on which the current paper builds, and Vistarini (2019). Vistarini discusses moduli spaces in string compactifications, for other purposes than the ones we pursue in this paper.} 
Similarly, the role of parameter spaces in scientific theories is seldom discussed. This stands in stark contrast to the contemporary physics literature, where, for good reasons that I hope to illustrate, these notions take centre stage.

In this paper I wish to point out that, in many areas of physics, there is a vast literature with precise studies of these issues. In short, my central thesis has two components, which are further specifications and consequences of the two general points that I labelled (i) and (ii) above (these points will be developed in Sections \ref{oneup} and \ref{structurem}, respectively):

(1)~~{\it Lifting `theory' and `model' ``one level up''.} Advances in physics, particularly the discovery of dualities, prompt us to be flexible about the use of the words `theory' and `model'. Specifically, we will be primarily concerned with how to formulate a `theory of theories'. What was formerly called `theory' is now called `model' so that a model will be (a) general and not a specific solution and (b) exact and not approximate---so lacking common connotations of `model' in the (vast!) modelling literature. Thus according to the idea of generality discussed above, we lift our usage of `theory' and `model' ``one level up''.\footnote{De Haro (2017a:~p.~261) and De Haro and Butterfield (2018:~p.~309) already proposed that dualities prompt moving one's usage of `model' and `theory' one level up. In this paper, I defend this idea in contexts where the models are not duals, and also combine it with the second thesis, namely that there is additional structure on the set of models that is now the theory, one level ``up''.}

Contrast this with the literature on theoretical equivalence cited above, which typically focusses on formal criteria for {\it comparing} different theories, rather than {\it expanding} the notion of a `theory' to encompass former theories as its models. Furthermore, while dualities are the main reason for the ``up''-move in physics, dualities are not often considered in discussions of theoretical equivalence.\footnote{A commendable exception is Weatherall (2019, 2020). This is of course not to deny that theoretical equivalence has been a main topic of discussion in the recent philosophical literature on dualities: for an overview of this literature, see De Haro and Butterfield (2025:~pp.~93--98) and De Haro (2021).}
Hence my earlier qualification that the literature on theoretical equivalence addresses point (i) only `in part'.

(2)~~{\it Structure on the set of models.} My proposal will be that one should not view a theory as a collection of models, but rather as a single geometric object. Thus one equips one's set of models successively with topological, differential, algebraic and geometric structure.\footnote{`Successively' here marks a conceptual and mathematical procedure: starting from a set of models, one adds, in order of dependency, topological, differential, algebraic, and geometric structure. This mirrors standard mathematical practice (and the chain of forgetful functors from geometry down to set theory) and is intended to make explicit the kinds of structure that occur in contemporary physics. The ordering does not reflect a commitment to ontological priority of e.g.~sets over topology.}
For example, there is a norm from which one can define a metric: and often also other geometric structure, such as complex and symplectic structure. Since one wishes such structure to admit a physical interpretation, it is natural to take it from contemporary physics rather than from purely logical or mathematical considerations. It is the absence of such algebraic-geometric structures in discussions of theoretical equivalence, together with additional structures that the next Section will review in detail, that prompts my previous qualification: the literature addresses point (ii) `not at all'.\footnote{Concrete proposals for equipping the set of theories with structure include those by Halvorson and Tsementzis (2017), mentioned earlier. They propose to discuss the ``universe'' of scientific theories as a two-category of categories. In addition, topological structure is discussed in Halvorson (2019:~p.~79ff). However, it is fair to say that, on balance, their work remains in the realm of logic, and does not consider structures of the algebraic-geometric type that I will discuss in this paper, which are required to describe contemporary physical theories. This comment is bolstered by their stated ambition to describe a ``universe'' of theories. For in the complex realm of physical theories, it seems best to start by describing a single ``planetary system'' in detail before attempting to describe a whole ``universe''. This is not a criticism, but a matter of their prioritizing ``toy'' theories expressed in a formal logical language.\label{planet}}

My claim is not that there is a single type of geometric object which, across {\it all} fields of physics, is what we call a `theory'. For I believe that, in the current state of development of physics, physical theories may be too heterogeneous for that. Indeed, such a claim would seem to require a reformulation of all of known physical theories. For example, there are important differences between theories with a finite, or with an infinite, number of degrees of freedom. Furthermore, flexibility as to different types of formulations that a theory can have is a welcome feature. Instead, my claim is that a theory is a {\it single structured object}: even though its type can vary from one area of physics to another. Having said that, two natural algebraic-geometric structures will appear in the examples: in the finite-dimensional case, differentiable manifolds (or cousins and generalizations like algebraic varieties) and bundles over them; in the infinite-dimensional case, normed spaces. 

Therefore we must answer the following central question: {\it Where in physics does the structure on the space of theories originate?} 
As my examples will show, this structure on the space of theories is determined chiefly by inter-theoretic relations such as (quasi-)\allowbreak dualities. In addition, some of the structure is induced from the structure of the configuration space or phase space (whether of finite systems or infinite ones: ``particles'' or ``fields''), and the dynamics and quantities thereon, i.e.~from the action or the free energy (depending on the types of examples). 

Such algebraic-geometric structures on theory space are not envisaged even by the combined view. This is for two reasons. First, what the combined view, as standardly understood in philosophy of physics, calls `theories' are, by our lights, models.\footnote{Note that it is not a requirement that models must be the solutions of the equations of a theory whose parameters are fixed. Logicians in fact often take theories to be specified much more generally, so that in particular the values of parameters need not be fixed.}
Second, and as I explained above, it is difficult to see where, within model theory, in the ``Tarskian'' sense used in logic and semantics, such geometric structures like a metric on moduli space or a symplectic form on the set of models could come from. Thus a new name is called for: I will call the view, that a theory is an algebraic-geometric object at the appropriate level of generality, {\it the geometric view of theories}.\footnote{It will be understood that the structures are not only topological and geometric, but also algebraic. Thus I take the term `geometric view' to cover all these aspects.}

The paper is organized as follows. Section \ref{GV} introduces the elements that suggest the need for a geometric view of theories. Section \ref{mproposal} then gives a specific proposal for the geometric view: namely, that a physical theory is a model bundlle. Section \ref{conclusion} is the conclusion. Two appendices include examples: Appendix \ref{SWth} discusses, as the main case study of this paper, the Seiberg-Witten theory, i.e.~the low-energy limit of ${\cal N}=2$ supersymmetric Yang-Mills theory. This case study illustrates the geometric view for a finite-dimensional theory. Appendix \ref{QC} briefly discusses another example, in quantum cosmology, where the theory space is of infinite dimension. 

\section{The geometric view of theories}\label{GV}

This Section will introduce materials, from both recent philosophy and physics, that will be the main ingredients of the general geometric view as here proposed. Indeed, the geometric view is motivated by the ideas of dualities and their generalizations, namely quasi-dualities, and the idea of the `parameter space' or `moduli space' of a theory. These align with the two  components of my central thesis about scientific theories, that I labelled (1) and (2) in Section \ref{intro}. For point (1), about lifting the usage of `theory' and `model' ``one level up'', is a natural consequence of the schematization of dualities. And point (2) is closely related to the idea of a quasi-duality and other structures, like moduli spaces, that appear in physical theories. Section \ref{oneup} first discusses the notion of duality, so as to explain point (1). Section \ref{structurem} discusses the notions of quasi-duality and moduli spaces, so as to introduce point (2). Section \ref{ssm} discusses the notions of model and of duality that will be used in Section \ref{mproposal} to give a more specific proposal for the geometric view.

\subsection{Dualities: lifting the usage of `theory' ``one level up''}\label{oneup}

A {\it duality} is an isomorphism between physical theories. This isomorphism is defined with respect to the structure that the duals share: a duality obtains {\it only} when that common structure is itself a bare theory (so that not any shared structure yields a duality). By a `bare theory', I mean basic structure that comprises a physical theory, regardless of interpretation: here, I follow De Haro and Butterfield (2018) in taking this structure to be a set of states or state-space ${\cal S}$, a set (almost always: an algebra) of quantities ${\cal Q}$, and a dynamics ${\cal D}$, together with a rule for assigning values to quantities on states. Thus a bare theory is a triple $T=\bra{\cal S},{\cal Q}, {\cal D}\ket$, and under a duality the set of values for the quantities also match between the duals.\footnote{The state-space and the set of quantities almost always come equipped with structures, for example symmetries. The duality then also needs to preserve these structures. For a discussion, see De Haro and Butterfield (2025:~pp.~7, 52, 60). Also, De Haro and Butterfield stress that the triple as here discussed is one of the ``shapes'' that a bare theory or model can take. The dynamics can be either deterministic, as in classical mechanics, or probabilistic, as in statistical mechanics. There are alternatives, in terms of path integrals, to the formulation in terms of states, quantities and dynamics. For a discussion, see De Haro, Teh and Butterfield (2017:~p.~75).} 
This shared structure $T=\bra{\cal S},{\cal Q}, {\cal D}\ket$, together with its assignments of values, is called the {\it common core}. The structure that duals do not share is called {\it specific structure}.\footnote{The distinction between common core and specific structure is a formal i.e.~non-interpretative one, because the notion of duality is formal i.e.~non-interpretative. Thus `specific' does not have the connotation of `redundant'. For example, there are cases where the specific structure should be interpreted as being non-redundant, such as the temperature in the Kramers--Wannier duality of the Ising model. For a discussion of this, and the corresponding distinction between external and internal interpretations, see De Haro and Butterfield (2025:~pp.~50--51) and De Haro (2017a:~pp.~261--264).}

An example of a duality is position-momentum duality in standard quantum mechanics, i.e.~the Fourier transformation between the position and momentum representations. The common core theory is the (infinite-dimensional, separable) Hilbert space $L^2(\mathbb{R}^3)$, equipped with an algebra of linear operators that is independent of the representation. The two duals are the momentum and position representations of both the Hilbert space and the algebra of operators (more precisely, a choice of a spectral family of projectors). Specific structure comes in because, in one representation, we have a real variable, ${\bf x}\in\mathbb{R}^3$, that denotes position, while in the other representation we have another real variable, ${\bf p}\in\mathbb{R}^3$, which denotes momentum. 

It is important to clarify that, from the point of view of the definition of a duality, this specific structure is not a matter of {\it interpreting} the variables in a domain of application, as being either `position' or `momentum': but rather a matter of {\it mathematical representation}, i.e.~of choosing structure that is not specified in the common core. Given any representation of the Hilbert space $L^2(\mathbb{R}^3)$, and of its canonical commutation relations, that diagonalizes one of the operators, there is another representation that diagonalizes the canonically conjugate operator (or, more correctly in the context of operators with a continuous spectrum, we find the spectral decomposition of this operator, i.e.~we represent it in terms of the spectral family of projectors associated with this operator). In the Hilbert space formulation that is the common core, we do not need to commit to a specific representation of the operators in terms of real variables.

After this brief introduction of dualities, I return to the two central components of my thesis from Section \ref{intro}, labelled (1) and (2) (I will discuss (2) in Section \ref{structurem}):\\
\\
(1)~~{\it Lifting the usage ``one level up''.} Since two duals share a common core that is itself a physical theory, it is natural to reserve the word `theory' for this common core, and to call the duals `models'. Thus the theory (here, the common core theory) ``stands above'' the dual models, which are mathematical {\it representations} or realizations of this theory. This is in effect point (1) from Section \ref{intro}, and it prompts the proposal of De Haro (2017a) to lift the usage of these words ``one level up''. (For more details, see De Haro and Butterfield, 2018, 2025.)

Although this choice for what we call `theory' and `model' may initially appear to be a mere choice of words: if combined with point (2), about structure, it is substantive.\footnote{De Haro (2017a:~pp.~266--270) already discussed that the choice is substantive once it is combined with an interpretation, which can be internal or external.} 
For it will lead us to a new formulation of the type of structure that we use to formulate scientific theories, and to a cautious realism about this structure (see Section \ref{mproposal}).

Part of the motivation to lift not only the word `theory', but also the word `model', one level up, is to keep the relation between `theory' and `model' fixed, so that a theory is more abstract and general: while the models, being mathematical representations of the (common core) theory, are concrete realizations specified by adding to the common core their specific structure.

Dualities between theories can be very unexpected, and so some amount of caution and flexibility in what one calls `theory' and `model' is called for. In one of the examples discussed in Appendix \ref{QC}, the dynamics of geometries in an expanding universe is mapped to the motion of a point particle in a moduli space. More generally, we will see that quantum cosmology at late-times in an expanding universe is dual to an object that is a sum over different partition functions in conformal field theory. On the combined view, one might have thought that different quantum field theories are independent of each other, and one might not have expected them to be parts of a single object. (In the original semantic conception, theories were seen as an unstructured collection of models: as discussed in Section \ref{intro}, only very recently has the idea of a structured set of models been given serious consideration by philosophers.) Yet this is precisely what quantum cosmology, and other examples from physics, force us to consider. In order to conceptualise this situation, which is not visible in examples from classical physics, we need to adopt a geometric view of theories. 

\subsection{Structure on the set of models: quasi-dualities and moduli spaces}\label{structurem} 

This Section first introduces the notions of quasi-duality and moduli space, which I will use to give a general characterization of the structure of contemporary physical theories, thereby developing point (2) of Section \ref{intro}. (Section \ref{mproposal} will make a more detailed proposal, based on the examples in the Appendix.)

A {\it quasi-duality} is a map between models that falls short of being a duality, i.e.~it falls short of being an isomorphism that preserves a common core theory. This can happen in either of the two ways in which a map can fall short of satisfying this definition: namely, through its failing to be a bijection, and through its failing to preserve a common core that is a theory. For example, it sometimes happens that a quasi-duality preserves {\it some} structure that the models have in common, but not enough, so that this common structure does not amount to a theory, i.e.~a triple $T=\bra{\cal S},{\cal Q}, {\cal D}\ket$.

An important type of quasi-duality in physics is an {\it effective duality}: it maps models that are approximate duals in a regime of parameters (for example, at large distances or low energies). Effective dualities sometimes, but not always, have dualities as limiting cases, when a parameter is taken to a limit, and other approximations and idealizations are made. For example, many of the ``dualities'' in string theory and quantum field theory are of this type: they are not cases of exact isomorphism, but they approximate an isomorphism within a specified range of parameters.\footnote{In the philosophical literature, effective dualities have been discussed in De Haro (2017b, 2019a) and De Haro and Butterfield (2025:~Sections 3.4, 7.5, 9.1, 9.3, 10.2, and Chapter 14). Recently, Eskens (2025) has used the concepts of approximate isomorphism and limiting operations to set out a precise definition and a taxonomy of effective dualities, and illustrates them with examples.}

T-duality in string theory is an important example of an effective duality. It is a map between momentum and winding states of a string, i.e.~it maps a string that is moving on a circle of radius $R$, with momentum $p=n/R$ and zero winding, where $n$ is an integer, to a string that has a winding number $w=n/R$ around the circle, and zero momentum. Note that such a duality is only possible for an extended object, like a string, that can wind around a circle.\footnote{For more on T-duality, see Read (2016:~pp.~216--218), Huggett (2017), De Haro and Butterfield (2025:~pp.~287--292), and Cinti and De Haro (2025).} 
T-duality is a duality only in the limit that the interactions between strings are small. By analogy with coupling constants in quantum field theory, interactions between strings are small when the string coupling parameter, $g_{\sm s}$, is small. This means that T-duality is a duality in the limit $g_{\sm s}\rightarrow 0$, but not if the string coupling is large.\footnote{The condition for T-duality to be valid is that the radius of the eleventh dimension, $R_{11}=g_{\sm s}\ell_{\sm s}$, where $\ell_{\sm s}$ is the string length, is small compared with the typical length scale that the string probes in a given state. The condition that the string coupling goes to zero is only a sufficient condition for the parameters: T-duality is also a duality if the string coupling is not strictly zero, so long as the string length $\ell_{\sm s}$ goes to zero so that $R_{11}$ is negligible compared to the other length scales that the string probes.}
Thus since T-duality is a duality only for a limited range of the theory's parameters, it is an effective duality.\\
\\
{\it Theory parameters and theory space.} How general, in terms of its adjustable parameters, should a theory be? In this question, I mean `theory' in my ``one-level-up" sense. For example: is general relativity, for slightly different values of Newton's constant, the same, or a different, theory? Although there may be some intuitive appeal to the ``specific'' answer, viz.~that a theory with the same form but different parameters is a different theory, I will argue that there is a real advantage to allowing the parameters to vary, thus viewing a theory as something more general. For, comparing models to each other at various values of the parameters within the larger ``theory space'', i.e.~within the framework of a single theory, gives us information that we would not find if we only consider fixed values of the parameters.\footnote{Jacobs (2023b:~p.~814) defends the view that the parameter spaces of constants such as Newton's constant $G$ are linked to spacetime, so that one should `conceive of Newtonian gravitation as a theory set on space-time-mass'. Jacobs's (2023b:~pp.~810--811) account is committed to anti-quidditism, because including $G$ within the theory's structure makes talk about quiddities unnecessary. According to anti-quidditism, physical quantities have no primitive identity, but (by analogy with anti-haecceitism) are qualitatively individuated by their relations to other masses.} 
For example, the metric on this space of theories contains physical information that we could not find only by looking at individual models. Instead, this information can be found by considering a class of models all sharing the same values of parameters. This realizes the idea of a `theory of theories', and it is part of the sense in which the notion of `theory' may be lifted ``one level up''.\\
\\
{\it Moduli space as theory space.} In mathematics, especially in the context of differential and algebraic geometry, the notion of a parameter space is made precise as a {\it moduli space}: moduli are discrete or continuously variable parameters that characterize a class of objects, for example a class of spaces. These parameters form a moduli space, whose geometric properties give us information about the class of objects.\footnote{It is important to not confuse the moduli space, which is usually a differentiable manifold made of points, with the model spaces (usually $\mathbb{R}^n$ or $\mathbb{C}^n$) where the coordinates of these points take values. The physical moduli (such as coupling, temperature etc.) take values in these model spaces, which set the ranges of the physical parameters.}
Namely, objects that are of similar type are in each other's neighbourhood in the moduli space. Alternatively, we can informally think of the neighbourhood of a point on the moduli space as all the possible ways in which, by changing the parameters by a small amount, a given object can be deformed to an object of the same type.\footnote{This way of thinking about moduli is in Kodaira (1986:~p.~228). Kodaira's notion of `completeness' gives a condition for the moduli space to encode all the (infinitesimal) deformations of a given object within a family of objects, as a surjective map between the set of objects and the moduli space (pp.~230, 284). See also Vistarini (2019:~p.~109).}
Thus moduli spaces are equipped with a topology, and often also a metric, which formalizes the notion of `closeness'.

A first example of a moduli space is the moduli space of isomorphic compact {\it Riemann surfaces of genus $g$} (where the `genus' is the number of handles, i.e.~holes, of a two-dimensional surface).\footnote{The criterion of isomorphism that is relevant for Riemann surfaces as discussed here is biholomorphism, i.e.~a bijective holomorphic map whose inverse is also holomorphic, i.e.~complex differentiable, which implies that it depends on the complex coordinate $z$ but not its complex conjugate. This means that isomorphic Riemann surfaces are identical in terms of the complex structure defined on them.}
It is the space of metrics on a compact Riemann surface, modded out by the group of diffeomorphisms that are continuously connected to the identity, and by the group of Weyl rescalings (Nelson, 1987:~p.~348). For $g>1$, the complex dimension of the moduli space is $d=3g-3$.\footnote{Riemann (1857) introduced in German the term `Modul', with plural `Moduln', and argued that a Riemann surface has $3g-3$ moduli. For more on this moduli space, see Giacchetto and Lewanski (2024:~pp.~4--8). For some mathematical details, including the construction of the moduli space of the torus and the calculation of the dimension of the moduli space of a Riemann surface, see Schlichenmaier (2007:~pp.~71--79). For the general theory of Riemann surfaces, see Farkas and Kra (1980:~Chapter III).} 
(For $g=0$, i.e.~a two-sphere, there are no moduli, and $d=0$.) For a genus-one surface, $g=1$ i.e.~a two-torus, $d=1$. The modulus is the complex structure $\t$ of the torus, which determines the ratio of the two periods of the torus in the complex plane: namely, the period of the B-cycle of the torus divided by the period of the A-cycle. It does so through the identification $z\cong z+n+m\tau$, where $z$ is a complex coordinate on the torus, $n,m\in\mathbb{Z}$ are integers that determine the two cycles, and $\tau$ determines the relative shape of the cycles, and whether they have any twists. Thus by varying the complex structure $\tau$ in its moduli space, we get tori whose cycles have all possible ratios. The moduli space of the complex torus is ${\cal M}=\mathbb{H}/\mbox{PSL}(2,\mathbb{Z})$, where $\mathbb{H}$ is the upper-half plane and $\mbox{PSL}(2,\mathbb{Z})$ is the group of projective special linear $2\times2$ matrices with integer entries, which acts on $\t$ as the group of fractional linear transformations with integer entries and unit determinant.\footnote{Although $\t$ takes values in the upper-half plane $\mathbb{H}$, the identification of the parameter $\t$ under the action of $\mbox{PSL}(2,\mathbb{Z})$-transformations in effect reduces its range to what is called the `fundamental domain': namely, an infinite strip of unit width above the unit circle in the complex place (Blumenhagen et al., 2013:~pp.~136--137). This is because if the A- and B- cycles of two tori have the same mutual ratio (even though they may have different sizes), or are related by a twist, then the tori are biholomorphic and thus conformally equivalent. Thus the moduli space ${\cal M}$ is a space of {\it isomorphism classes} of biholomorphic tori.  Each class is labelled by a unique point in ${\cal M}$, with a coordinate $\t$ taking values in the fundamental domain of the upper-half plane. The rest of the upper-half plane corresponds to the images of this fundamental domain under the action of $\mbox{PSL}(2,\mathbb{Z})$ on $\t$. Conversely, any $\t\in\mathbb{H}$ can be mapped into the fundamental domain by a $\mbox{PSL}(2,\mathbb{Z})$ transformation. For more details about tori, see Kodaira (1986:~pp.~48--49).\label{bitori}} 
Although this moduli space is smooth, in general, the moduli space is not a manifold, but an algebraic variety, i.e.~roughly speaking, it can be obtained as the solution of an algebraic equation (it is not a manifold because it can be ``pinched'' at certain points).\footnote{See Belavin and Knizhnik (1986:~p.~202). A moduli space is a connected, smooth, complex orbifold, i.e.~it is locally the quotient of a complex Euclidean space by a finite group $G$, so that it looks locally like an open set of $\mathbb{C}^d/G$.} 
The idea of `closeness' in the moduli space of Riemann surfaces is that a small deformation of the complex structure of a Riemann surface corresponds to a short path in the moduli space.\footnote{In the philosophical literature, the significance of the moduli space of Riemann surfaces for string theory and string dualities has been discussed by Rickles (2011).}

As a second example, take another space of metrics: namely, the {\it space of Einstein metrics} on a compact smooth manifold up to the diffeomorphisms of the manifold that are continuously connected to the identity. Here, an Einstein metric is one whose Ricci tensor is, up to a constant, equal to the metric tensor (and we consider the case that the metric tensor has positive signature, i.e.~it is a Riemannian manifold). This constant can be positive, zero, or negative, where the constant being zero is the familiar case of vacuum general relativity. These classes of isometric Riemannian manifolds of constant curvature, also called `Einstein structures', form a finite-dimensional moduli space (Besse, 1980:~pp.~340, 358). In general, for a given manifold, its moduli space of Einstein structures is not a manifold (for example, some of these moduli spaces have singular points, or are disconnected), although it does have a covering that is a smooth manifold.\footnote{For some of the properties of the moduli space, like its having constant scalar curvature, its being Hausdorff, and its having isolated points that cannot be deformed in the way discussed above, so that it corresponds to a unique object that does not belong to a continuous family of related objects, see Besse (1980:~pp.~352--355). Examples of moduli spaces and local deformations are at pp.~365--366; see also Kodaira (1986:~pp.~Chapter 4).}

Moduli spaces, as just defined, are an essential ingredient of the geometric view, because they are part of the structure on the set of models: they are a sort of background that ``holds the models together''. Thus we can now motivate the second component of my central thesis in Section \ref{intro}. Recall that the general idea was to view a theory not as a collection of models, but as a single geometric object (a more specific proposal will follow in Section \ref{mproposal}):\\
\\
(2)~~{\it Structure on the set of models.} In the examples given in the Appendix, models do not come on their own. They come with moduli spaces that are equipped with geometric structure, such as a metric. This additional structure makes moduli spaces useful, since it gives us information about the relations between the objects that correspond to the points of the moduli space. This leads in to the general geometric view: {\it A physical theory is a single structured object: namely, a set of models and a moduli space for those models, with topological and geometric structure on it.} 

For example, the metric of the moduli space, which is in general different in an obvious way from a spacetime metric, lives on the tensor bundle of the moduli space, i.e.~it assigns a symmetric bilinear form on each tangent space. And, as we will see in the examples in Appendix \ref{SWth}, the models, each comprising a set of states and quantities, live on fibres over the moduli space. The main idea of fibre is a family of objects---for us: models---associated with (so imagined as ``standing above'') an element of the base-space---for us: the moduli space. Agreed, in the context of differential geometry, `fibre bundle' has a technical geometric meaning that includes e.g.~smoothness conditions. I will indeed be committed to this technical meaning, which recurs in my main example in Appendix \ref{SWth}. However, in other examples that are formulated using algebraic geometry, the base-space can be an algebraic variety. The analogous (more general) structure required is then a sheaf, which can be defined using algebraic properties, rather than smoothness. The model bundle is not a locally trivial fibre bundle also in cases where around special parameter values (e.g.~$\hbar=0$) there is no neighbourhood that admits a trivialization, and the fibre there is not isomorphic to a generic fibre (see footnote \ref{feintzeig} for an example.)
I will briefly return to such generalizations in Section \ref{gc}.

In view of the above discussion, it is natural to propose that in general, for a large class of interesting examples, we have a bundle whose sections are the values of quantities, varying over the moduli space. The central idea that leads to the geometric view of theories is to apply the notion of a moduli space as a space of parameters of some class of objects, to scientific theories. We do this {\it by taking the objects to be models}. This will be my main proposal in Section \ref{mproposal}: I will call the resulting bundle, with a moduli space in the base and a model (states and quantities) in the fibres, a {\it model bundle}.

Note that model bundles are very different to the one obvious place where fibre bundles appear in physics: namely, principal bundles as the mathematical backbone of gauge theories.\footnote{There are different conventions in the literature about the use of the words `fibre bundle' and `principal bundle', which however will not cause any confusion. Physics-oriented texts usually take the typical fibre of a principal bundle to be a Lie group (Isham 1999, p.~220; Nakahara, 2003, p.~363), while mathematically-oriented texts often characterise principal bundles via the group-action (Hamilton, 2017, p.~207). Also, physics-oriented texts often define a fibre bundle as having a structure group (Nakahara, 2003, p.~350), while other texts do not require this explicitly. Nothing hinges on these conventions, and I will adopt the physics convention of equipping a fibre bundle with a structure group.}
Principal bundles are used in the context of classical field theory to incorporate local gauge symmetries of fields. By putting the structure group in the fibre, the group elements in the fibre are allowed to vary over the spacetime, which gives rise to a larger group of local gauge transformations, namely the group of bundle automorphisms. In this way, principal bundles are helpful in making the redundancies of gauge theories manifest. By contrast, in a model bundle there are models in the fibre, i.e.~sets of states and quantities that have a physical interpretation, typically characterized by vector bundles:\footnote{For more on how vector bundles can characterize states and quantities, see Appendix \ref{sqvb}.} 
and there is a moduli space of parameters or fields in the base manifold. The structure group is the group of quasi-dualities, which in general are not redundancies but rather relate distinct physical situations. The contrast can be summed up as follows: principal bundles give us a way to handle redundancy in gauge theory, while model bundles give us a way to handle physical states and quantities that depend on parameters. 

In this bundle picture, moduli spaces will always be either manifolds, perhaps with singularities, or algebraic varieties (and then one expects to generalize the bundle to a sheaf). The whole bundle need not be a manifold or an algebraic variety. Also, we will see that a model bundle covers a large class of cases where the fibres are isomorphic to each other. In examples where the fibres are not isomorphic, these require a further generalization (see the discussion in Section \ref{gc}). In other words, my strategy is to build my way up from examples in physics, using the minimal mathematical level of generality required, and generalizing only as required by the examples, rather than on a priori mathematical grounds. Using an astronomical metaphor, my aim is to start by describing theories that are familiar ``planetary systems'', before attempting to describe the whole ``universe'' of possible theories (cf.~footnote \ref{planet}).

To be able to give my general proposal, apart from giving examples from physics, we need a final ingredient: namely, a formulation of our models. This is done in the next Section.

\subsection{Formulation of the models and dualities}\label{ssm}

By a `model', I again mean a triple, $M=\bra{\cal S}, {\cal Q}, {\cal D}\ket$, comprising a set of states ${\cal S}$, i.e.~a state-space, an algebra of quantities ${\cal Q}$, and a dynamics ${\cal D}$, together with a rule for assigning values to quantities on states. (Additionally, the set of states and quantities usually comes equipped with symmetries, as automorphisms of these sets.) Because of this smooth assignment of values, the algebra of quantities and the state space are each other's duals, in the mathematical, not the physical, sense. Thus, given a quantity $Q\in{\cal Q}$, a state $s\in{\cal S}$ assigns to it a (real or complex) value, i.e.~$s(Q)=\bra s,Q\ket$, where $\bra s,\bullet\ket$ is the rule for assigning values to quantities that is part of the definition of the model. Thus we have ${\cal S}={\cal Q}^*$, as well as ${\cal Q}={\cal S}^*$. For a duality, we take the state-space ${\cal S}$ to be the set of dynamically possible states, which are a subset of the kinematically possible states.\footnote{Most classical dualities are valid at the level of the ``off-shell'' action, i.e.~without using the equations of motion. Hence these dualities can be extended to all kinematical states. However, the element that is common to all dualities is that the dynamical states of duals are isomorphic, hence our restriction to those states.} 
Thus, in effect, we can think of a model as a pair of a state-space and its dual, together with an assignment of a subspace of the state-space.

A duality between models is then an isomorphism between the state-spaces of two such models, $d:{\cal S}_1\rightarrow{\cal S}_2$, in such a way that the dynamics is preserved. Since the duality is to preserve the values of quantities on the states, the duality acts on the quantities as the (mathematically dual) map, $d^*$.\footnote{The dual map $d^*$ is $d^{-1}$ acting from the right, rather than from the left.} 
Thus a duality is a pair of maps, $(d,d^*)$. 

Models in physics depend on pararameters such as coupling constants and expectation values of some select subset of quantities. This also follows from the ideas (1) and (2) in Section \ref{intro}. When I wish to emphasise this, I will denote the model as: $M=M(t)$, where $t$ collectively denotes these parameters (see Section \ref{mproposal}). As we discussed in Section \ref{structurem}, these parameters form a moduli space.\footnote{Read (2016:~pp.~228) discusses the idea of duals defined over a moduli space, and how they can be embedded in the parameter space of a successor theory. See also Rickles (2011:~p.~65) and De Haro and Butterfield (2025:~pp.~484--499).}

Recall that the group of quasi-dualities is the structure group of a model bundle with this moduli space as the base space. Quasi-dualities were defined only locally, so that the action of the structure group varies above the base. Here, we have defined a duality map $(d,d^*)$ as acting on the models regardless of the parameters $t$. As we will discuss in Section \ref{pMB}, in the model bundle picture this is implemented by having the map act vertically on the fibres and uniformly across the whole moduli space, so that the duality map is defined globally. Thus the set of duality maps will form the structure group of a bundle that satisfies the additional condition of being globally trivializable (see Section \ref{rd}). 

\section{A specific proposal: models in the fibres}\label{mproposal}

This Section makes a structural proposal that (a) brings the remarks about quasi-dualities, moduli spaces, and fibre bundles from previous Sections, into a single structure; and (b) generalizes the examples of the geometric view that are discussed in two Appendices, and which aim to illustrate the main aspects of the view. (At this point, the interested reader is encouraged to read these Appendices.) I will propose that, at its simplest, 
a physical theory is a fibre bundle constructed in a specific way, with states and quantities in the fibres, and a space of parameters or moduli in the base. I will call this a {\it model bundle}.

The proposal is modest: I will present it while keeping in mind two types of possible limitations. I will first mention these limitations and then discuss the second one (since the first one was already addressed in Section \ref{intro}). First, the proposal is schematic, and in general not mathematically detailed. (However, as I will discuss in Section \ref{rd}, if the moduli space is the moduli space of complex structures of manifolds, a developed mathematical theory {\it does} exist: namely, Kodaira-Spencer theory). Thus the proposal and its implications are to be fleshed out in examples, as the Appendix aims to do. Second, illustrating the proposal may involve some amount of idealization, depending on the examples. 

About the second limitation: there might well be more general mathematical structures that cover more examples. Indeed I will discuss possible generalizations in Section \ref{gc}. But one must begin somewhere: and model bundles are certainly the simplest structures that one can find to cover cases such as the Seiberg-Witten theory. Also this second aspect is quite analogous to dualities: De Haro and Butterfield (2025:~p.~8) propose their isomorphism criterion for model triples...
\begin{quote}\small
`...in an undogmatic spirit. We accept that there will be rough edges in matching it to physicists' usage of the word `duality', and that maybe it could be improved. For there may be more mathematically precise conceptions of what a duality is than the one that we will defend. But that said, we are sanguine. We will see that the Schema fares very well. For it will prove apt to describe both simple and advanced physics examples, to cast light on relevant philosophical questions about dualities, and to give undoubtedly reasonable judgments of theoretical equivalence.
\end{quote}

Having discussed how these two points do not stand in our way, I will now propose to use a fibre bundle structure to implement the two ideas, (1) and (2), stated in Section \ref{intro}. So the remaining task is straightforward: this is the content of Section \ref{pMB}. Section \ref{philoD} discusses further philosophical aspects of this proposal.

\subsection{The proposal: A Model Bundle} \label{pMB}

This Section develops my main proposal. Section \ref{dmb} first defines a model bundle and its relation to quasi-dualities. Section \ref{rd} then discusses how to recover dualities from a model bundle, and Section \ref{gc} discusses some possible generalizations of this construction that may be suitable for more complex examples.

\subsubsection{Definition of the model bundle and quasi-dualities}\label{dmb}

To define the model bundle, we first gather the threads from Section \ref{ssm}. Recall that a model is a triple $M=\bra{\cal S},{\cal Q},{\cal D}\ket$ of structured set of states, structured set of quantities (usually, an algebra) and a dynamics. I will think of a model as a Cartesian product, $M={\cal Q}\times {\cal S}$. Thus my notation will suppress the dynamics, because in most cases of interest the dynamics can be stated as a choice of a distinguished quantity in ${\cal Q}$ that is the Hamiltonian (or, alternatively, the Lagrangian). A duality then maps the Hamiltonian of one model to the Hamiltonian of the dual model. I will assume that this condition is satisfied, and will not need to indicate it in my notation.\footnote{De Haro and Butterfield (2025:~p.~72) require that the duality map is equivariant for the two triples' dynamics (in either Schr\"odinger or Heisenberg pictures). For systems described by a Hamiltonian, this condition translates into the condition that the Hamiltonians match under the duality map. For example, in a quantum theory in the Schr\"odinger picture, the time evolution is implemented by the action of a unitary operator $U$ on states. The equivariance condition then comes down to the requirement that the time evolution operators of the two duals are related by the duality map $d_{\cal S}$ on states, as follows: $U_2=d_{\cal S}\,U_1\,d_{\cal S}^{-1}$.} 

Furthermore, recall from Section \ref{ssm} that a model depends on a set of parameters, i.e.~$M=M(t)$. The parameters take values in a moduli space, ${\cal M}$, which is in the base. It is natural for the moduli space to be the base, since the models depend on $t$.

To structure the mathematical aspects of the task, I will assume in this Section that quasi-dualities are local isomorphisms, i.e.~they are structure-preserving bijections above the neighbourhood of a point, even if they are not isomorphisms beyond this neighbourhood. (Recall that, in general, quasi-dualities need not be bijective. Thus this amounts to their ``failing to be a duality'' through the second way in Section \ref{structurem}, i.e.~not preserving a common core: I will return to the generalization of the current assumption in Section \ref{gc}.) Since they are bijections, we can take them to form a quasi-duality {\it group} $G$. In the Seiberg-Witten example, as well as in other examples of electric-magnetic dualities, $G=\mbox{SL}(2,\mathbb{Z})$ (for these other examples, see De Haro and Butterfield, 2025:~Sections 4.3, 5.3, 9.2, and Chapter 7).

By analogy with our earlier examples, the obvious suggestion is to then put the models in the fibres.\footnote{In the context of inter-theoretic reduction, Landsman (2017), Feintzeig (2020, 2022), and Steeger and Feintzeig (2021) define a {\it continuous bundle of C*-algebras} over a locally compact Hausdorff space, where the fibres are a family of C*-algebras, the base is a parameter space, and the continuous sections assign to each point in the base an element of the corresponding C*-algebra. Here, the parameter space is typically the set of values of Planck's constant $\hbar$ (in the case of quantum mechanics) and, in cases involving a system with $N$ degrees of freedom (with $N$ large), $\{1/N|N\in\mathbb{N}\}\cup\{0\}$. This illustrates how the idea of a model bundle can also be used to discuss inter-theoretic relations like reduction: in these works, the bundle construction is used to investigate the quantum-to-classical transition (i.e.~sending $\hbar\rightarrow0$) or a transition between a system with a finite and with an infinite number of degrees of freedom (e.g.~a quantum spin system, sending $N\rightarrow\infty$). Note that a continuous bundle of C*-algebras is not a fibre bundle, because no local trivialization exists at $\hbar=0$. 
See Landsman (2017:~pp.~294, 297) and Feintzeig (2020:~p.~621). \label{feintzeig}}
 
\begin{figure}
\begin{center}
\includegraphics[height=4.5cm]{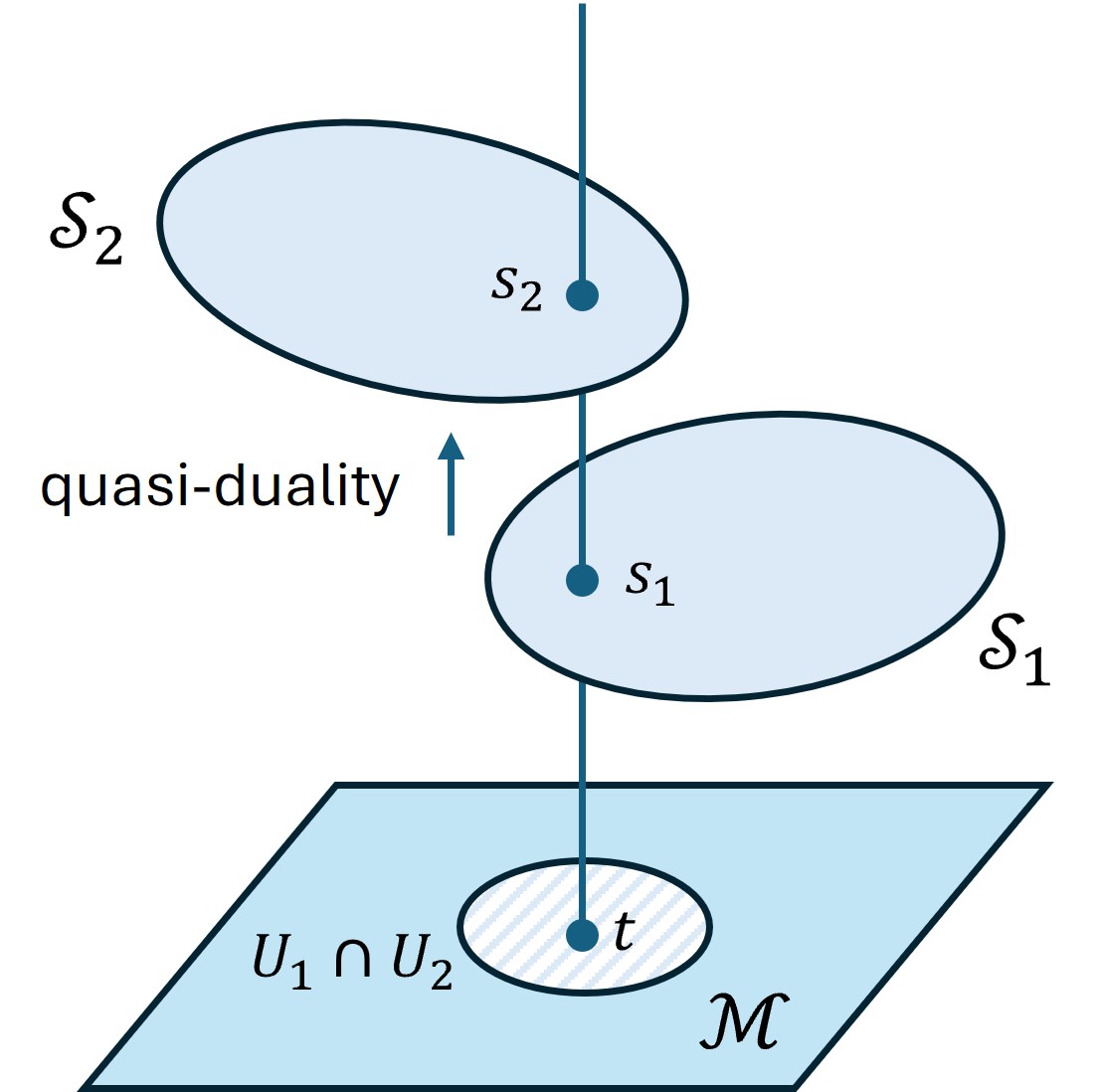}
\caption{\small A model bundle: quasi-dualities act vertically on the fibres above overlaps between regions of the moduli space. No given coordinatization needs to cover the whole bundle.}
\label{modelbundle}
\end{center}
\end{figure}

Thus we define a bundle $F\rightarrow E\overset{\pi}{\rightarrow}{\cal M}$, with total space $E$, fibres $F$, and projection $\pi:E\rightarrow{\cal M}$, i.e.~${\cal Q}(t)\times {\cal S}(t)\mapsto t\in{\cal M}$, that is a differentiable surjection, so that ``nothing in the base manifold is left out''. If there are quasi-dualities, as I will in general assume, we put models that are related by a quasi-duality on the same fibre. As I will now explain, these quasi-dual models will be different elements of the fibre over a given point that are related by a diffeomorphism of the fibre. This is natural, because the quasi-dualities form a group that we can identify with the bundle's structure group $G$, namely the group of fibre automorphisms, i.e.~the diffeomorphisms of the fibre that preserve the structure of the fibre. This group acts vertically on the fibres and maps the models in the fibres to each other. More precisely, there are bundle charts, $\phi_U:E_U\rightarrow U\times F$, where $E_U\subset E$ and $U\subset {\cal M}$, that satisfy projection and consistency conditions. The consistency conditions, also called `cocycle conditions', defined on overlaps of up to three regions, secure that the local pieces of the bundle can be glued consistently. Since the bundle charts allow us to write the bundle locally as a Cartesian product of the base and the fibre, they locally trivialize the bundle. By combining a chart and another chart's inverse map, we can define transition functions
$\f_j\circ\f_i^{-1}|_{(U_i\cap U_j)\times F}:(U_i\cap U_j)\times F\rightarrow(U_i\cap U_j)\times F$. These transition functions are local diffeomorphisms of the bundle induced by the local mapping of the bundle into the Cartesian product. Note that, by the projection condition on the bundle charts, these transition functions act only vertically on the fibre, i.e.~they leave the point $t\in U_i\cap U_j$ in the base fixed. This implies that we can use them to define a diffeomorphism that acts purely on the fibre above a point and leaves the base space fixed, i.e.~its input and output are only the fibre. We do this by evaluating the transition function on a given point $t\in U_i\cap U_j$, and using a subscript in our notation: $\f_{it}^{-1}:F\rightarrow E_{U_i}$, defined as $\f^{-1}_{it}(f):=\f_i^{-1}(t,f)$, where $(t,f)\in U_i\times F$. In this way we get, above every point $t\in U_i\cap U_j$ of the base, a diffeomorphism of the fibre: $\f_{jt}\circ\f_{it}^{-1}:F\rightarrow F$. These diffeomorphisms stay on a single fibre: they are elements of the structure group $G$ that acts vertically, which are the quasi-dualities. They contain the global information about the fibre: namely, they specify how the models in the fibre are glued. It is important to note that they are only defined above points of the base that lie in the intersection (overlap) of two regions of the kind $U$ that parametrize the set of charts, i.e.~$t\in U_i\cap U_j\subset{\cal M}$.\footnote{The fact that these diffeomorphisms are defined only locally, i.e.~on overlapping regions, can also be emphasised as follows. The diffeomorphisms allow us to define maps from the basis into the structure group $G$ of diffeomorphisms of the fibre, $\phi_{ji}:U_i\cap U_j\rightarrow G$, given by $t\mapsto\f_{jt}\circ\f_{it}^{-1}$, where $t\in U_i\cap U_j$. These maps are also called transition functions (of the fibre).}
For an illustration of a model bundle, see Figure \ref{modelbundle}.\\
\\
{\bf Weak and strong coupling in $S$-duality.} One might think that this statement---quasi-dualities act locally, on the fibre above a point---contradicts the physics of dualities, where a model at strong coupling is often related to one at weak coupling. The classic example is electric-magnetic duality, which takes the coupling parameter $\t$ to its inverse, i.e.~$\t\mapsto-1/\t$. In other words, quasi-dualities relate models at different values of the coupling (e.g.~this is so in Seiberg-Witten theory)---and this one might think means relating models in different fibres. However, one should here distinguish between the {\it parameter} that parametrizes the points of the moduli space, viz.~$t\in{\cal M}$, and its {\it value}, i.e.~$\t(t)$, which differs in different coordinate systems. $\t$ is the modulus of the torus, so that the transformation $\t\mapsto\t'=-1/\t$ is the $S$-generator of the modular group $\mbox{SL}(2,\mathbb{Z})$, which generates an isomorphic torus (see Section \ref{structurem}), and leaves the point $t$ invariant on ${\cal M}$, i.e.~it maps $\t(t)\mapsto\t'(t)$. In other words, $\t$ and $\t'$ are different coordinates for the {\it same point} $t\in{\cal M}$.\footnote{Recall, from Section \ref{structurem}, that the complex structure $\t$ was introduced in a coordinate-dependent way, through the identification $z\cong z+n+m\tau$, where $z$ is a complex coordinate on the torus. In other words, the parameter $\t$, which takes values in the upper-half plane $\mathbb{H}$, is a {\it coordinate} on the moduli space: see footnote \ref{bitori}.}
This implies that, in this type of duality, an inversion of the coupling constant does not move us horizontally to a different point of the base with a different fibre above it, but is a vertical automorphism of the bundle.\footnote{This can be seen explicitly in the Seiberg-Witten theory, where the base manifold is not parametrized by a coupling parameter, but by the expectation value of a physical field, $u$. Although the variable in the base is indeed related to the complex coupling, i.e.~$\t=\t(u)$, this relationship is complicated. Since $\mbox{SL}(2,\mathbb{Z})$ is the quasi-duality group that acts {\it vertically} on the fibre, $\tau$ is better thought of as a characterisation of  the fibre of the vector bundle: namely, $\tau(u)=a_D(u)/a(u)$ is the ratio of the coordinates in the fibre, on which the structure group $\mbox{SL}(2,\mathbb{Z})$ acts.\label{Fvertical}}
In other words, the physics of $S$-duality is entirely consistent with the way I have described it using the model bundle. Of course, this does not imply anything about the duals' being, at different values of the coupling, either theoretically equivalent or inequivalent: that is an additional interpretative question. I am here only saying that the dual models at different couplings, because they are related by the action of the structure group, are above each other in the same fibre.


The bundle charts defined above give us an interesting way to think about the elements of the fibre, i.e.~the quasi-dual models $M_i(t)$, as different coordinatizations of the fibre $F_t$ over a point $t\in{\cal M}$. Each coordinatization is given by a transition function evaluated over the point $t$, i.e.~$\f_{it}$, which takes values in $F$. More precisely, the model $M_i(t)\subseteq F$ is the image of $\f_{it}$, with as input all those points of the bundle, $e\in E_{U_i}$, whose projection is $t$, i.e.~$\pi(e)=t\in U_i$. Thinking of models as coordinatizations of the fibre, which is itself coordinate-independent (as is the whole bundle), agrees with the contrast in De Haro and Butterfield (2025:~p.~7) between a common core theory and its dual models. In that contrast, the models are representations (in the mathematical sense) of the common core theory, each model adding specific mathematical structure. Thus here, the fibre is the local common core (i.e.~for a fixed value of the moduli) and its elements are models, which are coordinatizations of this {\it local common core theory}. This also justifies a way of speaking in physics, where it is often said that dualities and quasi-dualities map models that express the local common core theory in `different coordinates'. This is literally the case for a model bundle, because the local trivializations of the bundle give us choices of coordinates where the quasi-dualities between models are changes of coordinates.\footnote{As I emphasised in Section \ref{structurem}-(2), this way of speaking uses the additional differential structure of the fibre bundle, as against a bundle or a sheaf. In other examples, algebraic rather than differential structure may be required: see footnote \ref{sheaves}.}

It is then clear that the fibre $F_t$ over a point is the local common core theory $T$, i.e.~{\it for a particular value of $t$}. (This important point will provide a crucial insight to recover dualities, and thus the whole common core, below, by requiring an additional condition.) For each model $M_i(t)$ is a representation of this fibre: namely, a coordinatization of it, given by $\f_{it}$.  This means that, when there is no duality but there is a quasi-duality group, the fibre bundle gives a generalization of the notion of a common core.

Is there a mathematical theory of such bundles? I mentioned above that I am not aware of such a {\it general} theory. However, in the case that the base space is the moduli space of complex structures of a complex manifold, there {\it is} fortunately such a general mathematical theory: namely, the Kodaira-Spencer theory of complex structure deformations (Kodaira, 1986:~Chapter 4). This theory describes differentiable families of compact complex manifolds, and how they vary as the complex structure $t$ changes. One centrepiece of this theory is that infinitesimal deformations are generated by vector fields that can be thought of as derivatives of the manifold in the fibre with respect to the complex structure $t$. These vector fields are cocycles, i.e.~closed forms (derived from a cocycle condition like the one we discussed in Section \ref{dmb}) that determine a cohomology group that describes the infinitesimal deformations.

The Seiberg-Witten bundle discussed in Appendix \ref{SWth} is of this type, since the moduli space point $t$, which in a local coordinate system $u(t)$ is the expectation value of the Higgs field (and written $u(t):=\bra\Tr\phi^2\ket$) can indeed be interpreted as the complex structure of the associated Riemann surface (even if it is not the usual complex structure $\tau$ on the torus that we discussed in Section \ref{dmb}). Thus that construction generalizes.\footnote{Vistarini (2019:~100-101) has studied complex structure deformations in connection with the question of the background-independence of string compactifications. With important differences from the geometric view that I propose here, she proposes to consider Hilbert spaces fibred over a moduli space of parameters. In my formulation in terms of a model bundle, this corresponds to the case where the state-space is a Hilbert space. Furthermore, one should also consider the algebra of quantities in the Hilbert space to be in the fibre. However, quasi-dualities do not seem to play any role in Vistarini's work, while for me they are the essential piece that gives the bundle its non-trivial structure. Likewise, Vistarini follows Vafa (1998:~p.~540) in envisaging dualities as isomorphisms between different bundles, while I recover a duality as a global isomorphism of models in the fibre of a given bundle (of course, the equivalence of different model bundles is a natural question once a model bundle has been constructed). There are other differences between Vistarini's (2019) treatment and mine that are worth noting. First, her main interest is in the question of background-independence in the context of string theory compactifications, while mine is in giving a geometric view that can be applied to many other physical theories, regardless of gravity. Furthermore, although Vistarini also considers state-spaces (specifically, Hilbert spaces) on fibres, she takes each fibre to come with a copy of $\mathbb{C}$ where the quantities take values. In contrast, I take models of both state-spaces and algebras of quantities in the fibres, with coordinates that are maps to model spaces, given by the expectation values of operators.}

\subsubsection{Recovering dualities}\label{rd}

\begin{figure}
\begin{center}
\includegraphics[height=4.5cm]{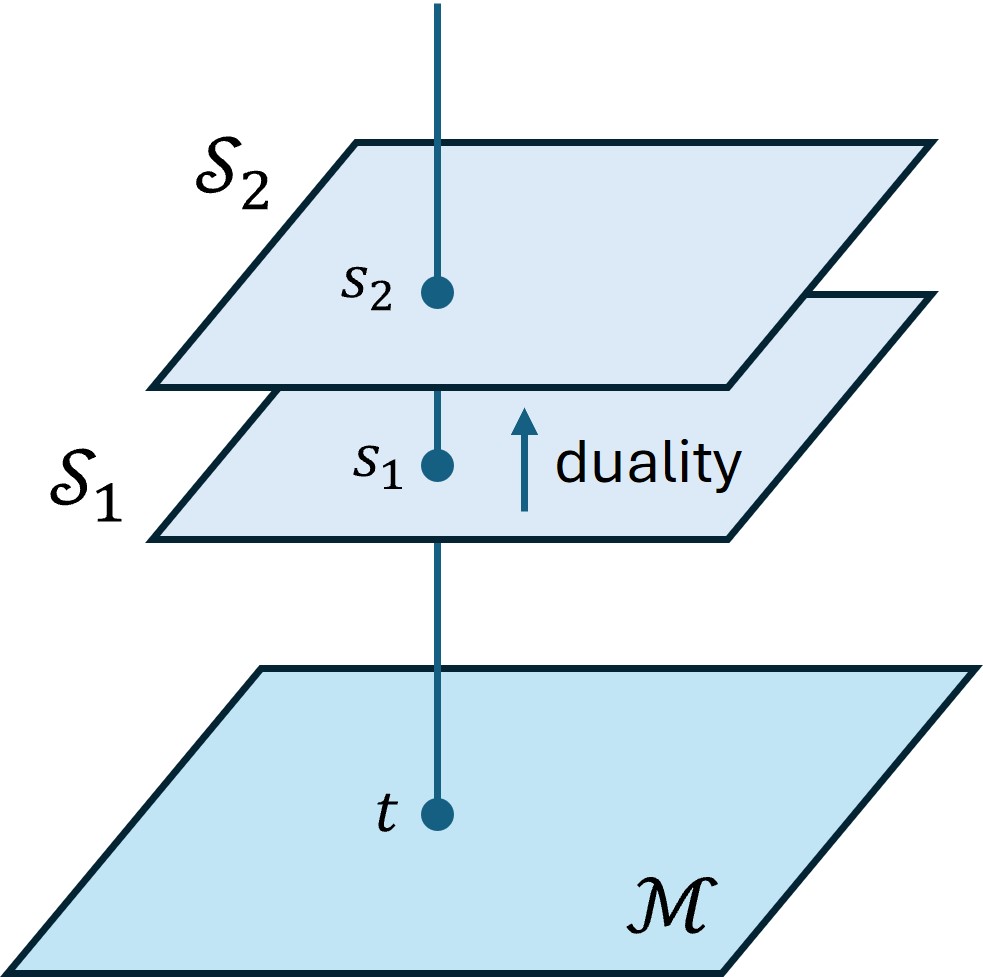}
\caption{\small If there is a duality, the bundle is trivial and the models are defined over the whole moduli space.}
\label{dualitybundle}
\end{center}
\end{figure}

In order to recover dual models as special cases of a model bundle, i.e.~to recover dualities, I first discuss the role of quasi-dualities in the model bundle. Recall that, since the structure group is in the codomain of the transition functions, these determine the quasi-dualities. Since the transition functions are defined on overlaps, quasi-dualities are local isomorphisms, i.e.~isomorphisms of two models that are both points in the fibre above a single given point $t$ in the moduli space ${\cal M}$. In particular, we discussed in Section \ref{dmb} that it is natural, in the context of differential geometry, to think of the models as giving different coordinatizations of the fibre, with quasi-dualities being local transition functions between such coordinatizations. And recall, from Section \ref{ssm}, that dualities are defined across the base, regardless of the value of the parameter $t$. In other words, in the language of differential geometry, a {\it duality} is a transition function that is defined over the whole moduli space. Thus the duality, as well as the models themselves, can be extended globally across the bundle. The models are well-defined over the whole moduli space only if the coordinates, given by the local trivializations, extend over the whole moduli space. This means that the bundle can be globally trivialized, and a single trivialization can in principle be used for the whole bundle. In other words, the validity of the models, which depends on the validity of the chosen coordinates, extends to the whole moduli space. This can also be stated in terms of the {\it common core}: there is a duality when the common core of the models extends to the whole moduli space: see Figure \ref{dualitybundle}. 

This can also be understood in the opposite direction: a map between models can fall short of being a duality, and thus be a quasi-duality, in two ways (see the beginning of Section \ref{structurem}): through its failing to be a bijection, or through its failing to preseve a common core that is a theory. We will briefly discuss non-invertible quasi-dualities, i.e.~quasi-dualities that fail to be bijections, in the next Section. The failure of the quasi-dualities to extend to the whole moduli space, that we discussed above, is indeed a case of the quasi-duality's not preserving a common core theory for all values of $t$: in other words, the quasi-duality is defined only for fixed or special values of $t$, but not throughout the whole moduli space. (Recall, from above, that the fact that some quasi-dualities relate models at different values of the coupling is consistent with the fact that both dualities and quasi-dualities act locally, on the fibre above a point. For the illustration of this fact in the Seiberg-Witten theory, see footnote \ref{Fvertical}.) 

Although the bundle is trivial if we have a duality, i.e.~it is the direct product of the base and the fibres, it is still useful to have different coordinatizations of it. For, as is well-known from studies of dualities, going to a dual formulation can e.g.~make a difficult problem tractable. For example, $S$-duality inverts the coupling $\t(t)$ to its dual value in a different coordinate system, viz.~$\t'(t)=-1/\t(t)$, thus turning a strong-coupling problem into a weak-coupling one (see the discussion of $S$-duality in Section \ref{dmb}).

\subsubsection{Generalizations of the construction}\label{gc}
 
As I discussed in the preamble of Section \ref{mproposal}, the model bundle proposal can (and perhaps should) be generalized, based on more complex examples. I will here briefly indicate the most obvious directions to explore, based on the structure we have found. However, as I mentioned before, I believe it is crucial to ground such generalizations on actual examples in physics that require us to generalize: rather than generalizing for the sake of mathematics, in a vacuum of actual physical examples. 

The model bundle's structure can be weakened in two obvious directions: `vertically' and `horizontally'. (i) In the vertical direction, we can weaken the action of the structure group $G$ by taking it not to be a diffeomorphism of the fibre. This means that the quasi-dualities are not local isomorphisms, i.e.~isomorphisms on the local fibres (recall that, in a model bundle, a quasi-duality is taken to be a local diffeomorphism). As usual, this can happen in either of two ways: (1) Through the quasi-duality's not being structure-preserving: this is the case in the Seiberg-Witten theory, where we simplified the theory by not taking into account its spin structure. Taking into account this spin structure would give a bijection that is not locally structure-preserving. (2) Through the quasi-duality's not being a local bijection. For example, the quasi-duality might be non-invertible, and then one would get a semi-group, rather than a group, $G$. Non-invertible quasi-dualities appear in recent discussions of the quantum Ising model.\footnote{A prominent example is non-invertible Kramers-Wannier quantum quasi-duality: see Seiberg and Shao (2023) and Shao (2023). Combining non-invertible quasi-dualities with a bundle structure results in a semi-group model bundle. I am not aware of major uses of these structures in physics.}
(ii) In the horizontal direction, we can weaken the assumption that there is a generic fibre $F$ to which the fibres over each point $t$ are isomorphic. Thus one might consider a bundle that is not a fibre bundle (see e.g.~Isham, 1999:~Section 5.1) or, more generally a sheaf.\footnote{One of the physical motivations for considering sheaves is the need to define vector bundles over submanifolds of a space (Sharpe 2003:~Section 2.3). At its simplest, a sheaf associates a mathematical object, such as a set, a group, a module or a category, to every open set on a topological space. Informally, the definition of a sheaf usually begins by associating sections, i.e.~the mathematical objects under consideration, to the open sets of the topological space, and then considering restriction maps on these sections under nesting of the open sets. To obtain a sheaf, additional conditions are required; see Eisenbund and Harris (2000:~pp.~11--12). In quantum field theory and string theory, certain D-brane states on Calabi-Yau manifolds are related to sheaves over a moduli space, which can be used to understand mirror symmetry, as an instance of the (geometric) Langlands correspondence. More precisely, the geometric Langlands correspondence is realized as an equivalence between: (i) certain electric branes, defined on the moduli space ${\cal M}(G,C)$ of flat holomorphic $G$-bundles on a compact Riemann surface $C$, which are solutions of the Hitchin equations with gauge group $G$ on $C$; and (ii) magnetic branes, which are described by a sheaf of differential operators on the moduli space with Langlands-dual group $^LG$ (for a simple definition of the Langlands-dual group, see De Haro and Butterfield, 2025, p.~243; Gukov and Witten 2006, pp.~168--175). The moduli space ${\cal M}(G,C)$ has singularities whose physical interpretation is similar to that in the Seiberg-Witten theory. In the neighbourhood of these singularities, the low-energy sigma models that are used to realize the Langlands correspondence are not valid, and one has to go ``up'' to four dimensions, to a version of ${\cal N}=4$ supersymmetric Yang-Mills theory. The Langlands correspondence originates in the $S$-duality of this ${\cal N}=4$ supersymmetric Yang-Mills theory. For discussions, see Kapustin and Witten (2007) and Kapustin (2008), who focus on the geometric Langlands programme; and Gaiotto and Witten (2022), for its analytic version. \label{sheaves}}

\subsection{The semantic conception of theories, dualities, and scientific realism}\label{philoD}

The semantic conception of theories, as originally formulated, does not envisage structures like the ones we have here discussed for the model bundle. For that reason, there is no answer within the semantic view to the question I asked at the end of the Introduction: {\it Where in physics does the structure on the space of theories originate?} Also the recent literature on theoretical equivalence, which sometimes does postulate some structures on the set of models, does not offer an answer to this question (recall Section \ref{intro}). Likewise, there are no answers to be found in the literature about e.g.~the realist interpretation of moduli spaces and the metrics on them. Therefore, to present a theory as a model bundle, a generalization of the semantic conception is required, and many of the basic philosophical questions, such as those regarding theoretical equivalence and scientific realism, are again ``on the table''. Formally, the semantic view is then recovered as the trivial model bundle, which is globally the product of the fibre and the base. Furthermore, the semantic view usually does not envisage quasi-dualities, so that the structure group is the identity. Thus the semantic view is the special case of a trivial model bundle, which has no particularly interesting structure. In this Section, I will make some comments about both these formal and interpretative issues. A more detailed discussion is left for future work.

As I discussed in the previous Section, the case of a duality, as against quasi-duality, is the case where the transition function is defined over the whole moduli space. A duality in effect gives a different writing of the fibre over the base. The explicit inclusion of the base in this formulation emphasises the fact that dualities often depend on a set of parameters, i.e.~a moduli space.

Since we recover dualities when the transition functions are globally defined, the model bundle itself can be seen as a generalization, to quasi-dualities rather than dualities, of the common core. It is a common core that is obtained not by abstraction, i.e.~by ``deleting the specific structure not shared by the models'', but by augmentation, i.e.~by moving to a larger theory that incorporates the models including their specific structure, and relates them through partial isomorphisms.\footnote{For a discussion of these two ways to get a common core theory, see De Haro and Butterfield (2025:~Section 12.2.1).} 
This is like the idea of sophistication about symmetries.\footnote{For more on this discussion, see for example Dewar (2019) and the reply by Martens and Read (2021).}
In other words, the model bundle includes the dual models, and realizes the quasi-dualities through different coordinatizations of the bundle (this last aspect is also true for dualities, where the coordinatizations extend through the whole base).

Thus we return to our main question {\it Where in physics does the structure on the space of theories originate?} In the model bundle, the key structure comes from the way the models are tied together in fibres above overlapping regions, and this information is provided by quasi-dualities. In the Seiberg-Witten theory, quasi-dualities allow us to focus on the states that can be probed at low energies above a given point in the moduli space. Thus the validity of the low-energy description in a given region of the moduli space determines the branch of the fibre that we are in. In the rest of this Section, I will mostly use the example of the Seiberg-Witten theory to answer the above question.

These effects, as well as the non-triviality of the metric, come from quantum effects on the theory's quantities, especially in terms of the theory's Lagrangian. In the Seiberg-Witten theory, the non-triviality of the metric is the result of the infinite series of renormalization effects of the prepotential that is obtained from the low-energy Lagrangian.\footnote{In classical gauge theories, moduli spaces are also equipped with metrics. For the original derivation of the moduli space metric, and how it can be obtained from the classical action, see Manton (1982:~p.~55). For more discussion, see Atiyah and Hitchin (1988:~pp.~12--13). A beautiful derivation of the metric and its Kahler structure for vortices is in Manton and Sutcliffe (2004:~pp.~205--212).} 
In quantum cosmology (see Appendix \ref{QC}), the metric comes from the evaluation of the wave function, which is expressible in terms of CFT partition functions that play the role of wave-functions equipped with a metric. 

As I have suggested before, the model bundle, as a formal structure, is not a substitute for substantive interpretative discussions of theoretical equivalence and dualities. Nothing in the model bundle guarantees that duals or quasi-duals are automatically theoretically equivalent models, and in general they are inequivalent, because the models' projections to the base are only partly overlapping. Thus the same semantic and epistemic requirements for theoretical equivalence that have been discussed in the recent literature, are also requirements for the theoretical equivalence of the fibres.\footnote{Here, I in particular have in mind the requirement of unextendability (De Haro, 2017a), and the interpretationalism-motivationalism contrast by Read and M\o ller-Nielsen (2020). For a discussion of necessary and sufficient conditions for theoretical equivalence of physical theories, see De Haro and Butterfield (2025:~Chapter 12). See also Rickles (2017:~pp.~63--66) and Huggett and W\"uthrich (2025:~Chapter 8).}
For this reason, the interpretative issues that are interesting here do not mainly concern fibre bundle realism:\footnote{For example, the present discussion is separate from the question whether physical degrees of freedom are ``really'' in a principal bundle or only in a vector bundle. For more on that discussion, see Jacobs (2023a:~pp.~39--40) and Gomes (2025:~pp.~512--514).} 
for presenting a theory as a model bundle is not advancing a na\"ive or `direct' model bundle realist view.\footnote{See North (2021:~pp.~4, 218), whose direct realism is distinguished from a na\"ive reading of the physics directly from the mathematics.}
Rather, the realism must be cautious, in that substantive logico-semantic analysis is required before ontological conclusions can be drawn from a particular theory formulation. Thus the catious realist takes inter-theoretic relations seriously in addressing ontological questions, both within the model bundle itself (intra-theoretical equivalence), as between different model bundles (inter-theoretical equivalence).\footnote{For a more detailed discussion of this cautious realist attitude, see De Haro and Butterfield (2025: Section 13.2). Other discussions of realism in the context of dualities include Rickles (2013:~p.~319) and Dawid (2013).}
Therefore, even though I defend no direct realism about model bundles, lifting the notions of `model' and `theory' one level up and putting models together into a single object like a model bundle is a substantive issue for the cautious realist. It is not merely a structural way of comparing models, but it has ontological consequences. 

In particular, there is a natural realism about the geometric structures on the moduli space that I here endorse in the examples. This is because the geometric structures such as the metric and its set of singularities arise from the physics of the problem. In the Seiberg-Witten example, the non-trivial metric comes from the kinetic term in the low-energy Wilsonian effective action for the fields, and so it describes their propagation speeds and energies. A change in the metric, for example a rescaling as we move through the moduli space, changes the propagators of the Seiberg-Witten theory by a corresponding scale factor (which is inversely related to the coefficient of the kinetic term). This will change the effective mass of the low-energy particles, which means that the relation between their speed and energy changes. Other changes in the metric modify the structure of the light-cones as ``seen'' by these particles or low-energy fields, and thus the causal structure of the effective spacetime that these fields explore. Geodesics change, and particle excitations follow different trajectories, compared to those in Minkowski spacetime. Thus, although the background spacetime is flat, the metric on the moduli space has effects that resemble those of a metric in spacetime. Note that the Seiberg-Witten metric is not the spacetime metric of the Wilsonian effective action, but rather a metric on the space of fields, analogous to the background or ambient spacetime metric in string theory.\footnote{See Seiberg and Witten (1994a:~p.~32). For a philosophical exposition, see Vergouwen and De Haro (2025:~p.~7).} 
Nevertheless, the effects on the motions of particles are similar, in terms of the changes to the geodesics and speeds of particles that result. 

Also, the global properties of the metric carry important physical information. The monodromies of sections around singularities contain information about massless states that emerge unexpectedly in the low-energy spectrum. They are `unexpected', because these states are not part of the point-particle excitation spectrum that one finds at weak coupling, but they are non-perturbative states.\footnote{For the sense in which these states emerge, see Vergouwen and De Haro (2025).}

In the UV completion of the Seiberg-Witten theory in string theory, the moduli space metric has a more direct realist interpretation: not just as the effective metric ``seen'' by the fields, but as induced from a higher-dimensional ambient spacetime metric. This involves a full geometrization of the theory, where the role of the moduli space metric is akin to the role of the ambient spacetime in string theory. The moduli space and its metric are encoded in the geometry of the higher-dimensional spacetime on which D-branes move and intersect, in particular via the Seiberg-Witten elliptic curve embedded in the brane worldvolume and the periods $(a,a_D)$ defined on it. Without entering into details: the ${\cal N}=2$ supersymmetric Yang-Mills theory whose low-energy limit is the Seiberg-Witten theory gets realized, at higher energies, as the world-volume theory on a set of fivebranes (with five dimensions of space and one of time) intersecting with fourbranes (with four dimensions of space and one of time) along three common spatial directions and time. This common $(3+1)$-dimensional spacetime is the spacetime of the ${\cal N}=2$ supersymmetric Yang-Mills theory. The moduli space of the Seiberg-Witten theory is then parametrized by displacements in two of the additional spatial dimensions of the fivebranes.\footnote{For details, see Witten (1997:~pp.~454--455), Giveon and Kutasov (1999:~pp.~1020--1021), and De Boer and De Haro (2004:~pp.~178--179). These papers also discuss the embedding of this D-brane configuration into M-theory, in terms of a single M5-brane. Note that, in the type IIA string realization that I discuss in the main text, the fourbranes are extended in one additional spatial dimension. This is the dimension along which they are suspended between two fivebranes (so that the fivebranes are located at fixed points in this dimension).}
The Seiberg-Witten metric is then obtained as the metric induced on these moduli (equivalently, on the deformations of the embedded elliptic curve) from the spacetime metric.\footnote{This realist attitude towards the moduli space metric contrasts with e.g.~Halvorson (2019:~p.~278), who mentions `certain ``nearness'' relations' between models. His use of scare quotes around ``nearness'' seems to indicate that distance relations on a space of models are not to be interpreted realistically.}

\section{Conclusion}\label{conclusion}

This paper has argued that physical theories are best conceptualised as structured algebraic-geometric objects. It has proposed model bundles as a framework for dealing with various theories in physics. This framework generalizes, and therefore subsumes, the semantic conception of theories, which is recovered as a special case: namely, a trivial bundle, i.e.~a product bundle, usually with a structure group that is the unity. Thus the semantic view, as usually conceived, is not rich enough to accommodate for the structures that we find in physics, such as a metric on the moduli space, and which have physical content. 

My proposal for the geometric view does not arise from an enthusiasm for any specific mathematical formalism. Rather, the view is based on examples from scientific practice, and its usefulness is illustrated by the way it helps us to systematise, and thereby gain understanding of, concrete examples. Having said that, the geometric structures here proposed are natural candidates for a realist interpretation---where such realism is again vindicated by the physical origin and interpretation of the geometric structures in the examples. 

In the finite-dimensional case, there is a manifold or an algebraic variety---a moduli space---with topological and geometric structure as the hallmark of the geometric view. In the infinite-dimensional case, the state space is directly equipped with a metric. In both cases, what were formerly `theories' are models of this larger theoretical structure. The examples, available in the physical literature for philosophical exploration, are numerous.\footnote{In addition to the examples given in footnote \ref{sheaves}, I here give an incomplete list of examples. A major example is the moduli space of soliton solutions and its relation to dynamics: see Manton and Sutcliffe (2004:~pp.~102--108, 202--205). More specifically, the moduli space of monopoles is discussed in Atiyah and Hitchin (1988:~pp.~14--20). The moduli space of instantons is discussed in Naber (2011:~pp.~370--376). Seiberg (1995) has discussed the moduli space of ${\cal N}=1$ supersymmetric gauge theories, which also enjoy a form of electric-magnetic duality.} 

The geometric view puts quasi-dualities, as against dualities, into focus, and it highlights their important role. This role of quasi-dualities goes beyond their being approximations of dualities, or their being cases of partial isomorphism. In the geometric view, quasi-dualities are transition functions on limited regions of the moduli space. As such, the geometric view highlights this previously under-appreciated role of quasi-dualities. Furthermore, it underlines how inter-theoretic relations can change our view of theories. 

In the usual conceptions of scientific theories, there is no clear preference for choosing theories as having either fixed or free parameters: or, in other words, for the level of generality that one chooses one's theory to have. In this sense, philosophical discussions are usually ``blind'' to parameters, which have no distinguished role within the theory. By constrast, the geometric view, through its ``lifting models one level up'' and its being geometrically unified, vindicates theories as being equipped with a moduli space of parameters. De Haro and Butterfield (2025:~pp.~29--30) have called this an `intermediate' notion of theory. Parameters are not simply variable real or complex numbers: they are geometrically organised in a space whose geometric structures have physical significance.

The geometric view opens up several additional questions, of which I will here mention three. First, the global information that it encodes is itself a natural candidate for a realist interpretation, and it is related to the emergence of new states. Thus an interesting open question is to characterise emergence in terms of the topological properties of the model bundle.\footnote{For a framework for emergence that allows for this kind of treament, see De Haro (2019:~pp.~17--19) and De Haro and Butterfield (2025:~pp.~506--507).} 
Second, developing a geometric view of theories seems closely tied to understanding issues in quantum field theory and quantum gravity. This is illustrated by the prominence of dualities and quasi-dualities in quantum field theory and quantum gravity research over the past several decades.\footnote{For more discussion of this point, see the discussion of successor theories in De Haro and Butterfield (2025:~Section 14.1).} 
Thus the geometric view can help theory development both formally, heuristically and interpretatively. Finally, the geometric view more generally seems closely related to the unifying role of dualities that has been stressed by Dawid (2013) in the context of string theory. Thus the geometric view is closely linked, and bound to cast light on, more specific issues in string theory, such as the string theory landscape (itself an example of the geometric view), compactification, background-independence, and the ontology of string theory.

\section*{Acknowledgements}
\addcontentsline{toc}{section}{Acknowledgements}

I thank Jeremy Butterfield, Enrico Cinti, and James Read for discussions and for their detailed comments on this paper. I also thank Silvester Borsboom, Benjamin Feintzeig, and Nick Huggett for discussions of some of these materials. I thank St John's College, Oxford, for their hospitality and support during the completion of this paper, in the summer of 2025. Finally, I thank the audiences at the places where, since 2023, this work has been presented: Oxford and Milan (2023), Utrecht and Nijmegen (2024), and Gd\'ansk and Stockholm (2025).

\begin{appendices}

\section{The Seiberg-Witten Theory}\label{SWth}

This Appendix presents the Seiberg-Witten (1994a,b) theory as the paper's main case study of the geometric view.\footnote{This is due to space constraints: Appendix \ref{QC} will briefly discuss another example. More examples are given in De Haro and Butterfield (2025) and Cinti and De Haro (2025).}
In the philosophical literature, the Seiberg-Witten theory has been discussed previously, from the point of view of emergence, in Vergouwen and De Haro (2025): and, from the {\it general} point of view of the geometric view, in De Haro and Butterfield (2025). Here, I will focus on a number of aspects that have not appeared in these works, and that motivate my specific proposal in Section \ref{mproposal}: namely, the theory's fibre bundle structure. Section \ref{introSW} first introduces the theory, emphasising the monodromies of quantitites. Section \ref{mgv} then uses this discussion to motivate the geometric view. Section \ref{gstb} discusses the geometric structure on the tangent bundle to the moduli space. Section \ref{sqvb} discusses the models in the fibres. 

\subsection{A survey of the Seiberg-Witten theory}\label{introSW}

The Seiberg-Witten theory is a fully quantum theory that describes the low-energy regime of ${\cal N}=2$ SU(2) supersymmetric Yang-Mills theory. The states of this theory, in the phase where the global SU(2) gauge symmetry is broken down to U(1) by the Higgs mechanism, are the states of a U(1) gauge field and the Higgs field (here, a complex scalar field) and their supersymmetric partners. In addition, there are non-perturbative solitonic states of e.g.~monopoles that I will discuss in a moment.\footnote{The physics jargon for this phase is `Coulomb branch', because the global gauge symmetry is only partly broken by the Higgs mechanism. The massive gauge bosons have been integrated out at low energies, and the only gauge excitations remaining are those of a massless U(1) gauge field, so that the effective potential between charges is Coulomb-like. For a discussion of the different branches of gauge theories, see De Haro and Butterfield (2025:~pp.~222--223). The branches of the Seiberg-Witten theory are discussed at pp.~269--270.}

The states are labelled by their electric and magnetic charges, $n_e$ and $n_m$ respectively (and their corresponding spins, to be discussed below). In addition, the Higgs field takes an expectation value on the states that, together with the charges, contributes to the masses. The mass $M$ of a state with charges $(n_m,n_e)$ satisfies the following equality:\footnote{States whose mass equals their charge are called BPS states, for Bogomol'nyi, Prasad and Sommerfeld.}
\bea
M=\sqrt{2}\,|Z|\,,
\eea
where $Z=an_e+a_Dn_m$ is called the {\it central charge}, in the sense that it gives a central extension of the super-Poincar\'e algebra and contains key dynamical information. $a$ parametrizes the degeneracy of the Higgs field, i.e.~the freedom that remains after solving its low-energy equation. This leads to a moduli space of vacua. This degeneracy also appears in the expectation value of the Higgs field, $\bra\f\ket$. $a_D$ is its magnetic dual, and it can be found locally by doing a Legendre transformation of the low-energy Wilsonian effective action of the ${\cal N}=2$ supersymmetric Yang-Mills theory.  

The central charge and thus the mass has an $\mbox{SL}(2,\mathbb{Z})$ symmetry that is a central topic in the Seiberg-Witten theory. Write the central charge as follows:
\bea
Z=n_e\,a_D+n_e\,a=(n_m~~n_e)\left(\begin{array}{c}a_D\\a\end{array}\right).
\eea
This pairing is obviously invariant under the following left- and right-matrix multiplication by $M\in\mbox{SL}(2,\mathbb{Z})$:
\bea\label{MMinv}
\left(\begin{array}{c}a_D\\a\end{array}\right)&\mapsto&\left(\begin{array}{c}a_D'\\a'\end{array}\right):=M\left(\begin{array}{c}a_D\\a\end{array}\right)\nn
(n_m~~n_e)&\mapsto&(n_m'~~n_e'):=(n_m~~n_e)\,M^{-1}\,,
\eea
where the restriction of the special linear group to the field of integers $\mathbb{Z}$ follows from the fact that the charges $(n_m,n_e)$ are integers. 

The fact that the states $(n_m,n_e)$ and the quantities $(a_D,a)^T$ appear as each others' mathematical duals, and transform in complementary ways under $\mbox{SL}(2,\mathbb{Z})$, expresses the mathematical duality between states and quantities that we expect more generally. The value $Z$ is obtained by evaluating the quantity $(a_D,a)^T$ on a state with given electric and magnetic charges. This will be important when we set up the bundle structure in the next Section.

At first sight, the action of $\mbox{SL}(2,\mathbb{Z})$ looks like a duality, because it maps a state with charges $(n_m,n_e)$ to a state with different charges, $(n_m',n_e')$, such that important physical quantities such as the central charge and the mass are preserved. Thus it looks like an isomorphism of state spaces that preserves the quantities.\footnote{However, this action of $\mbox{SL}(2,\mathbb{Z})$ falls short of being a duality, for two reasons. First, the states are not simply determined by their charges, but also by their spin quantum numbers: and the spin quantum numbers do {\it not} match across the duality, because states with different charges also have different spins, and the theory's spectrum is not invariant under the exchange. (Since the study of the duality is not my main topic in this paper, in most of this Section, I will set this point aside, and return to it at the end of Section \ref{sqvb}.) For more detailed explanations of why ${\cal N}=2$ is not a duality in this sense, see De Haro and Butterfield (2025:~pp.~268--269).}

The second reason is the main issue in this paper. Namely, the group of transformations $\mbox{SL}(2,\mathbb{Z})$ does not ``globally'' relate dual descriptions, i.e.~different ways of representing the same set of states and quantities, but rather expresses the {\it local} applicability of variables such as the charges and the expectation value of the Higgs field. In other words, as we vary the theory's parameters, our description becomes invalid,\footnote{Specifically, the free energy is an infinite series in powers of the expectation value of the Higgs field $a$, and this series diverges outside its region of validity. The change of variables from $a$ to $a_D$ is in effect a resummation of this series.} 
and it is necessary to do an $\mbox{SL}(2,\mathbb{Z})$ transformation so as to get a new description that is valid in that region of parameters. Thus the various descriptions are not equivalent representations of a common core defined for all values of the paramters, but rather representations with limited regions of validity overlapping only on partial regions of the space of parameters. In short, $\mbox{SL}(2,\mathbb{Z})$ is a group of {\it quasi-dualities}.

\subsection{Motivating the geometric view}\label{mgv}

The reason why the geometric view is needed is the second reason at the end of the previous Section: the states and quantities vary dynamically as we change the physical parameters, and there is no set of variables for the states or the quantities in the Seiberg-Witten theory that can cover the whole range of parameters. This Section fills in the details.\\
\\
{\it Moduli space of vacua.} As I mentioned above, $a$ parametrizes the (expectation value of) the Higgs field at the minimum of the potential, $\bra\f\ket$. The expectation value of the square of the Higgs field is denoted by $u:=\bra\Tr\,\f^2\ket$, and classically $u=\half a^2$, which means that the expectation value of the square of the field factorizes as the square of the correlation functions of that field (quantum mechanically, as we will discuss, this relation gets modified). Since the expectation value varies dynamically, there is a classical {\it moduli space of vacua}: namely, the space of states that, in the classical limit, satisfy the equations of motion. Although it might happen that there is a unique vacuum in the quantum theory, so that the degeneracy is removed, this is actually not the case in the Seiberg-Witten theory. The relation between the expectation value of the square of the Higgs, $u$, and the square of the parameter of the Higgs, $a^2$, gets one loop and non-perturbative corrections.\\
\\
{\it States, quantities, and their monodromies.} I will now discuss why the states and quantities are not simply given by, respectively, pairs of variables $(n_m,n_e)$ and $(a_D(u),a(u))$ that are well-defined everywhere on the moduli space (here, $u$ labels a point on the moduli space). We can show that, since the moduli space has singularities, $(a_D(u),a(u))$ are multi-valued: namely, they are local sections of a fibre bundle whose base is the moduli space. This follows from the following relations that are satisfied by the quantities for large values of $|u|$, i.e.~in the semi-classical region where the expectation value of the Higgs field is large:
\bea\label{asympt}
a_D&\simeq&{2i\over\pi}\,a\,\ln a\nn
a&\simeq&\sqrt{2u}\,.
\eea
These expressions have two noteworthy properties: (i) Due to the logarithm, the magnetic quantity $a_D$ is not single-valued and has non-trivial monodromy in the complex plane. Here, the monodromy is the effect of the function's not being single-valued when we go on a path that encircles the singularity (below, we will see that there is a monodromy matrix). (ii) Due to the square root, $a$ has non-trivial monodromy around $u=0$, and there is a branch cut from $(-\infty,0]$. Thus if we go on a counterclockwise contour of large radius (because $|u|$ is large) in the $u$-plane, $(\xi,u)\mapsto e^{\xi i}u$, where $\xi\in[0,2\pi]$, these two quantities change by:
\bea
a_D&\mapsto&a_D'=-a_D+2a\nn
a&\mapsto&a'=-a\,.
\eea
Going on a counterclockwise contour of large radius is homologically equivalent to having $u$ encircle the point $\infty$ counterclockwise. We can summarize this by introducing a monodromy matrix $M_\infty$:
\bea\label{monodromy}
\left(\begin{array}{c}a_D(u)\\a(u)\end{array}\right)\mapsto\left(\begin{array}{c}a_D'(u)\\a'(u)\end{array}\right)= M_\infty \left(\begin{array}{c}a_D(u)\\a(u)\end{array}\right);~~~~M_\infty=\left(\begin{array}{cc}-1&2\\0&-1\end{array}\right).
\eea

The states are dual to the quantities and transform, as in Eq.~\eq{MMinv}, with the inverse matrix:
\bea
(n_m~~n_e)&\mapsto&(n_m'~~n_e'):=(n_m~~n_e)\,M_\infty^{-1}\,;~~~~M_\infty^{-1}=\left(\begin{array}{cc}-1&-2\\0&-1\end{array}\right).
\eea
For example, a monopole has magnetic charge but no electric charge: $(n_m,n_e)=(1,0)$. The monodromy matrix maps this to a state with non-zero electric charge: it is a dyon, viz.~$(n_m',n_e')=(-1,-2)$. On the other hand, a purely electric state, $(n_m,n_e)=(0,1)$, remains purely electric, with charge of the opposite sign: $(n_m',n_e')=(0,-1)$.

This argument shows the $u$-dependence of the states: there are, above different points on the $u$-plane, different sets of states and quantities. Furthermore, due to the appearance of monodromies, the states and quantities are non-trivially fibred over the $u$-plane. This leads us to the geometric view of theories. 

The monodromy matrix in Eq.~\eq{monodromy} is related to the singularity at $u=\infty$.\footnote{For an explanation of why these are not merely coordinate singularities, see De Haro and Butterfield (2025:~p.~263). Their existence is related to the impossibility of the moduli space metric to be positive everywhere. Physically, as we will discuss below, the singular behaviour happens when the Wilsonian effective action breaks down, because certain states have not been taken into account. These states can then be studied through the study of the monodromies around the singularities.} 
There are two other singularities in the $u$-plane,\footnote{For an explanation of this statement, see Seiberg and Witten (1994a:~p.~36) and Bilal (1996:~pp.~106--107).} 
at points that are labelled $u=\L$ and $u=-\L$, where $\Lambda$ is a dynamically generated scale of strong coupling, where non-perturbative effects become significant. The associated monodromy matrices are:
\bea
M_\L=\left(\begin{array}{cc}1&0\\-2&1\end{array}\right);~~~~~~~~M_{-\L}=\left(\begin{array}{cc}-1&2\\-2&3\end{array}\right)~.
\eea
The monodromy at $u=\L$ gives, through Eq.~\eq{MMinv}, the following charges:
\bea
n_m'&=&n_m+2n_e\nn
n_e'&=&n_e\,.
\eea
We see that, compared with the monodromy around $u=\infty$, the roles of the electric and magnetic charges are interchanged at $u=\L$. Thus a pure electric charge, $(n_m,n_e)=(0,1)$, acquires {\it magnetic} charge and gets mapped to a dyon: $(n_m',n_e')=(2,1)$. A monopole state, $(n_m,n_e)=(1,0)$, remains invariant: $(n_m',n_e')=(1,0)$. At the third singularity, $u=-\L$, both monopoles and electrons are mapped to dyons. 

As explained in De Haro and Butterfield (2025:~p.~263), this can be interpreted in terms of perturbative and non-perturbative states becoming massless at the singularities. Recall that the Seiberg-Witten theory describes the low-energy regime of supersymmetric Yang-Mills theory. This means that massive states, whose mass is much higher than the dynamically generated scale $\L$, are integrated out and do not contribute to the low-energy spectrum. 

For example, in the weakly coupled regime near $u=\infty$, the monopoles of the ${\cal N}=2$ supersymmetric Yang-Mills theory, which are massive solitonic states, are integrated out, and so they are not part of the low-energy spectrum. If we did start with a monopole state some distance away from the singularity and took it around the singularity, thus crossing a branch cut in the complex plane, it would acquire electric charge and become a dyon. 

At the strong coupling singularity $u=\L$, monopoles become massless and appear in the low-energy spectrum. Their mass is given by $a_D(u=\L)=0$. This implies that taking an electron around the singularity, it acquires magnetic charge (it does not become a dyon, but rather a pair of particles: see Seiberg and Witten, 1994a, p.~39). At the third singularity, $u=-\L$, it is the dyons that become massless, because both electric and magnetic charges map to dyons. Which particle becomes massless at the singularity follows from the analysis of the eigenvectors of the monodromies. 

The three monodromy matrices $M_\infty, M_\L$ and $M_{-\L}$ span a subgroup of $\mbox{SL}(2,\mathbb{Z})$ called $\G_0(4)$, but not the whole $\mbox{SL}(2,\mathbb{Z})$.\footnote{The subgroup $\G_0(4)$ consists of those matrices, $\left(\begin{array}{cc}a&b\\c&d\end{array}\right)\in\mbox{SL}(2,\mathbb{Z})$, such that $b=\mbox{0 mod 4}$.}
We will return to the relation between these two groups in the next Section.

Outside of the singularities and branch cuts, $a_D$ and $a$ depend only on $u$ and not $\bar u$: they are holomorphic functions. This will be incorporated below in the requirement that the bundle is holomorphic.\\
\\
{\it Physical interpretation of the dual variable.} One way to derive the basic duality transformation is by doing a Legendre transformation of the Wilsonian effective action, which in effect exchanges electric and magnetic descriptions. It turns out to be an $S$-duality transformation, where $S$ is one of the generators of the duality group $\mbox{SL}(2,\mathbb{Z})$.\footnote{For more on $S$-duality, see De Haro and Butterfield (2015:~Sections 4.3 and Chapter 7).} 
In the original Lagrangian, the action depends on $a$ both directly, as well as through a quantity ${\cal F}(a)$ called the prepotential. The prepotential appears as a multiplicative factor in the kinetic terms of the scalar and gauge fields in the Wilsonian effective action. The Legendre transform in effect introduces an independent coordinate $a_D$ through:
\bea\label{aD}
a_D={\pa{\cal F}\over\pa a}\,.
\eea
This is analogous to switching between the Lagrangian and Hamiltonian formalisms. The prepotential, in effect, describes the gauge coupling and all of its quantum corrections, i.e.~a one-loop effect and an infinite series of instanton corrections: 
\bea\label{Fsum}
{\cal F}(a)=\half\t_0\,a^2+{i\over\pi}\,a^2\ln{a^2\over\L^2}+{a^2\over2\pi i}\sum_{k=1}^\infty c_k\left({\L\over a}\right)^{4k}.
\eea
Here, $\t_0$ is the bare coupling, and the remaining terms represent its quantum corrections. For more discussion, see De Haro and Butterfield (2025:~Section 7.5).

\subsection{Geometric structures on the tangent bundle}\label{gstb}

As I suggested in the previous Section, since the states and quantities are not single-valued functions on the moduli space of vacua, we should think of the structure of the Seiberg-Witten theory as a {\it fibre bundle} over the moduli space. Describing this fibre bundle will be the job of the next Section. This Section first remarks that the moduli space in the base of this fibre bundle is not a bare manifold: it comes with a tangent bundle on which geometric structures are defined.

As the previous Section discussed, the moduli space is the complex plane with the three singular points deleted, i.e.~${\cal M}=\mathbb{C}\backslash \{\L,-\L,\infty\}$. This is the space of low-energy vacua of the ${\cal N}=2$ SU(2) supersymmetric Yang-Mills theory, and it is parametrized by the complex variable $u$, which is the expectation value of the square of the Higgs field. At each point $p$, we have a tangent space $T_p{\cal M}$ of tangent vectors $V=v(p)(\pa/\pa u)|_p\in T_p{\cal M}$, and each $T_p{\cal M}$ is homeomorphic to $\mathbb{C}$. These tangent spaces are the fibres of the tangent bundle. As we will discuss below, the compatibility of the tangent bundle with the metric in effect restricts the structure group that acts on the fibres to be the subgroup $\mbox{SL}(2,\mathbb{R})$ of isometries of the metric.

In the Seiberg-Witten theory, the tangent bundle also comes equipped with a {\it Riemannian metric}, i.e.~a symmetric positive-definite bilinear form on the tangent space, $g_p:T_p{\cal M}\otimes T_p{\cal M}\rightarrow\mathbb{R}_{>0}$. 

The moduli space is equipped with a symplectic structure that is also metric compatible, i.e.~its values on tangent vectors are equal to the values of the metric, up to a factor of $i$, so that: $\forall v,w\in T_p{\cal M}$, $g_p(v,w)=\om_p(v,J_pw)$, where $\om$ is the symplectic structure and $J$ is the complex structure. The metric respects the complex structure, which means that it is Hermitian. Furthermore, the symplectic form is closed, $\dd\om=0$. Since it is also compatible with the complex structure $J$ and the Hermitian metric, it is a K\"ahler form, and the metric is K\"ahler.\footnote{For more on K\"ahler manifolds and K\"ahler metrics, see e.g.~Schlichenmaier (2007:~pp.~177--178) and Nakahara (2003:~pp.~330--332).}
This implies that the metric can locally be written as the total derivative of a K\"ahler potential, which is determined by the prepotential ${\cal F}$. This metric can be conveniently written as:
\bea\label{metric}
\dd s^2=\mbox{Im}\left({\cal F}''(a)\,{\pa a\over\pa u}{\pa\bar a\over\pa\bar u}\right)\dd u\otimes\dd\bar u\,.
\eea
It is not difficult to see physically where the metric comes from. As we discussed at the end of the previous Section, the second derivative of the prepotential appears as a multiplicative factor in the kinetic terms of both the scalar and gauge fields in the Wilsonian effective action, as a volume factor. This indicates that there is a non-trivial metric on the space of all the fields. When written in terms of the coordinates $u$ and $\bar u$ on the moduli space, one gets the metric Eq.~\eq{metric}.

In the next Section, I will discuss the vector bundle that is defined on the moduli space. Although, to streamline the discussion, I will not emphasise this, this vector bundle is not independent of the geometric structures that we have discussed in this Section. The two structures are closely related and satisfy compatibility conditions. For example, the fibres of the vector bundle come equipped with a symplectic form $\om=\mbox{Im}\,\dd a_D\wedge\dd a$ that, by considering the pullback of a map from the moduli space into the fibre, gives the symplectic structure on the moduli space discussed in this Section. Also, the same prepotential that gives the K\"ahler potential and thus the metric on the moduli space plays a crucial role in the fibres, since it relates the two coordinates of the fibre, $a_D$ and $a$, where $a_D$ is defined in Eq.~\eq{aD}. Finally, the group $\mbox{SL}(2,\mathbb{Z})$ is the discrete subgroup of the group of symmetries of the above K\"ahler metric, viz.~$\mbox{SL}(2,\mathbb{R})$. (The group $\mbox{SL}(2,\mathbb{Z})$ agrees with the group that we found through the analysis of the monodromy matrices in the previous Section. The fact that the group is discrete is a consequence of the quantization of the charges. Here, we have required that only this discrete subgroup of $\mbox{SL}(2,\mathbb{R})$ survives, i.e.~we have required the compatibility with the dual vector bundle.) However, $\mbox{SL}(2,\mathbb{Z})$ is not a gauge group but a duality group, and there is no gauge connection taking values in this group (in particular, the curvature is zero). Thus this vector bundle is {\it not} associated to a principal fibre bundle. 

\subsection{States and quantities on vector bundles}\label{sqvb}

In this Section, I describe the vector bundles where the states and quantities reside in the fibre. The main point of using a fibre bundle is that, as we move on the moduli space, the states and quantities change. Indeed, near the singularities new massless states appear, and the physical quantities like the mass and the total charge also change. This is described by monodromies, as in Eq.~\eq{monodromy}. We state how fibre bundles organize this information. 

We define a flat holomorphic vector bundle, $X\rightarrow V\overset{\pi}{\rightarrow}{\cal M}$, with general fibre $X$. Recall that, for our purposes, the quantities of interest are $a$ and $a_D$, other quantities being derived from these two. This means that the fibres are two-dimensional, locally $X\cong\mathbb{C}^2$, with the pair $(a_D,a)$ being a coordinate on the fibre. And as we also saw, the states $(n_m,n_e)$ are mathematical duals of the quantities, and so they live in the dual bundle. This means that the total bundle is the Whitney sum, or direct sum, bundle of the algebra of quantities and its dual bundle. Our discussion here will focus on the quantities. 

The projection $\pi:V\rightarrow{\cal M}$ is a surjective differentiable map (so that every element in the moduli space ${\cal M}$ has at least one inverse image in $V$). The base manifold of the bundle is the moduli space ${\cal M}$. That the bundle is {\it flat} means that it has no curvature: even if, as we are about to discuss, it does have non-trivial holonomies around singular points. The pairs $(a_D,a)\in X$ are coordinates in the fibres of the bundle. More precisely, the local differentiable function $u\in U\subset{\cal M}\mapsto(a_D(u),a(u))$ is a local holomorphic section $s:U\rightarrow V$ such that $\pi\,\circ\,s=\mbox{id}_U$. At any base point $u\in U$, one can find coordinates $(a_D(u),a(u))$ in the fibre above $u$, so that the pairs $(u;a_D,a)\in U\times X$ give a local trivialization of the bundle, i.e.~any point in a local neighbourhood of the total space of the bundle can be written in this way. 

The compatibility of the vector bundle with the tangent bundle structure defined in the previous Section (where I discussed compatibility in the last paragraph) requires that we restrict the group of bundle diffeomorphisms to the diffeomorphisms that fix the form of the metric in the base manifold, i.e.~the isometries of Eq.~\eq{metric}. This leaves us with the group $\mbox{SL}(2,\mathbb{R})$, which is the group of M\"obius (i.e.~fractional linear) transformations of the $u$-plane. Of these transformations, only the discrete subgroup $\mbox{SL}(2,\mathbb{Z})$, i.e.~the (double cover) of the group of modular transformations, are symmetries of the quantum theory. As we discussed, this is a consequence of the quantization condition for the charges.

This discrete subgroup induces the group of symmetries of the fibre, through the monodromies, discussed in Eq.~\eq{monodromy}, under $2\pi$-rotations around the singularities. Namely, the structure group that acts on the fibre is $\mbox{SL}(2,\mathbb{Z})$. It acts by $2\times2$ left-matrix multiplication as in Eq.~\eq{monodromy},\footnote{$\mbox{SL}(2,\mathbb{Z})$ is the modular group of fractional linear transformations, and it is useful to think of it as the group generated by the $2\times2$ matrices $S$ and $T$ that satisfy the relations $(ST)^3=\mathbbm{1}_{2\times2}$ and $S^4=\mathbbm{1}_{2\times2}$.
The duality group $\G_0(4)$ is generated by $T$ and $ST^2S^{-1}$ is a congruence group of $\mbox{SL}(2,\mathbb{Z})$, i.e.~it consists of matrices of $\mbox{SL}(2,\mathbb{Z})$ that satisfy additional congruence conditions. See Gukov and Witten (2006:~p.~58).}
so that taking an element in the fibre $(a_D,a)$ around a singularity in the $u$-plane in the base has the effect of multiplication by an element of the structure group.\\
\\
{\it Role of dualities and quasi-dualities.} Recall the second reason, at the end of the of Section \ref{introSW}, why $\mbox{SL}(2,\mathbb{Z})$ is not a duality, but rather a quasi-duality. The group of diffeomorphisms, which induces the structure group $\mbox{SL}(2,\mathbb{Z})$, acts only locally on overlapping regions in the base manifold. Thus the diffeomorphisms are not globally defined, and a given pair like $(a_D(u),a(u))$, written in these coordinates, is not valid everywhere in the moduli space. In particular, the singularities of $a_D(u)$ require an $\mbox{SL}(2,\mathbb{Z})$ transformation to a new pair of variables that are well-defined near the singular points of $a_D(u)$.

Physically, this stems from the fact that the sum in the prepotential, Eq.~\eq{Fsum}, has a limited radius of convergence in the $u$-plane: namely, it is only valid near $u=\infty$. For values of $u$ near $u=\L$, the series needs to be resummed. There is then an analogous sum written in terms of the dual variable $a_D$. Thus the prepotential is, just as $a_D$, a multi-valued section. 

Physics texts sometimes emphasise that the Seiberg-Witten theory is not invariant under S-duality, i.e.~the electric-magnetic duality given by the generator $S$ of the structure group $\mbox{SL}(2,\mathbb{Z})$ (see Lerche, 1998:~p.~181). This is because the generator $S$ is not part of the monodromy group, $\Gamma_0(2)\subset\mbox{SL}(2,\mathbb{Z})$, generated by the three monodromy matrices $M_1, M_{-1}$ and $M_\infty$. I endorse this, which aligns with my earlier statement that we only have quasi-dualities. However, it is important to note that we {\it do} use the $S$ generator in constructing the Seiberg-Witten theory: namely, it is the generator that takes us from $a$ to its dual variable $a_D$, and gives us the dual model, with its prepotential, that is valid near $u=\L$. This agrees with the fact that the structure group of the bundle really is the whole $\mbox{SL}(2,\mathbb{Z})$ and not just its subgroup generated by the monodromies. In other words, the additional generator is needed to make the prepotential finite in the regions where the semi-classical description breaks down.

\section{Quantum Cosmology and the WDW Equation}\label{QC}

My second example concerns theories of quantum cosmology: namely, pure quantum gravity in a spacetime with a positive cosmological constant. Due to space constraints, and having given a detailed example in Appendix \ref{SWth}, this example will get a briefer treatment. This example has two main features that make it of interest:

(i)~~The example being in quantum gravity, it shows that the geometric view extends to these kinds of theories.

(ii)~~Even though the example is not formulated in the language of fibre bundles, it will illustrate the general points of the geometric view, (1) and (2) in Section \ref{intro}. That is: (1) Structures that we normally call `theories', are now models. (2) There is geometric structure on this set of models (here, metric structure). Thus the theory is a normed space. 

About (i), it is worth noting that the geometric view is especially natural in theories of quantum gravity and string theories, which often have holographic reformulations. In these holographic reformulations, a spatial or temporal dimension gets mapped to the renormalization group scale of a quantum field theory, so that, at the level of the theory's formal structure, the difference between a `space of parameters' or `moduli space' and a real spacetime, disappears. 

\subsection{A wave-function is a CFT partition function: models ``one level up''}

In quantum gravity, diffeomorphism invariance imposes constraints on the wave-functional, the most important constraint being the Wheeler-De Witt (WDW) equation, 
\bea
{\cal H}\,\Psi[g]=0\,, 
\eea
where $\Psi[g]$ is the wave-functional depending on the three-metric $g_{ij}$ on a spatial slice, and ${\cal H}$ is the Hamiltonian constraint.\footnote{We consider here the situation with pure gravity, and so I do not include matter fields. Matter fields can be straightforwardly included: see Chakraborty et al.~(2024:~pp.~6, 14).} 
The WDW equation is highly non-linear, and so it is difficult to study. Traditionally, it has usually been studied in a mini-superspace approximation, where one assumes that the metric is homogeneous and isotropic, so that the spatial metric depends on a single time-dependent scale factor, and the configuration space is finite-dimensional (here, parametrized by the scale factor). However, in a recent series of papers it has been shown how to go beyond the mini-superspace approximation, by expanding the constraints near future infinity.\footnote{See Chakraborty et al.~(2023, 2024) and Godet (2024).} 
Since the universe is expanding, an expansion near future infinity is a large-volume approximation. This approximation is both cosmologically relevant, and gives important insights beyond the mini-superspace approximation.\footnote{One of these insights is the fact that the theory satisfies a version of holography: more specifically, it illustrates a version of the `principle of holography of information'. This is discussed in Chakraborty et al.~(2023).}

I will summarize the result and then state how it illustrates the geometric view. The main question is to find the space of wave-functions $\Psi[g]$ that satisfy the WDW equation. We write the wave-function as:
\bea
\Psi[g]=e^{iS[g]}Z[g]\,,
\eea
where $S[g]$ is a universal phase factor local in the metric. Chakraborty et al.~(2023) show that one can solve for these functionals in the large-volume approximation in asymptotically de Sitter spacetimes. $Z[g]$ is a non-local functional that satisfies an anomaly equation similar to the one in AdS-CFT.\footnote{However, unlike in AdS-CFT, it is such that its modulus-squared, $|Z[g]|^2$, is invariant under the diffeomorphism group and under Weyl rescalings.}
It can be expanded in powers of Newton's constant, $\k=8\pi G_{\tn{N}}$, as follows:
\bea
\ln Z[g]=\sum_n\k^n{\cal G}_n\,,
\eea
where the solutions are the following functionals:
\bea
{\cal G}_n={1\over n!}\int\dd y\,\dd z~G^{\{ij\}}_n(y,z)\,h_{i_1j_1}(z_1)\cdots h_{i_nj_n}(z_n)\,,
\eea
and the $h$'s are linear perturbations of the spatial metric. Here $z_1,\ldots,z_n$ are ``holographic'' coordinates at timelike future infinity, denoted collectively by $z$, that are integrated over. $\{ij\}$ denotes the (symmetrized) collections of indices $(i_1,\ldots,i_n)$ and $(j_1,\ldots,j_n)$, and the Einstein summation convention for repeated indices is used. For the technical details, see Chakraborty et al.~(2024:~pp.~20--21). The main point is that, near the Euclidean vacuum, this allows us to write the solution of the WDW equation in a similar way to the above expression, i.e.~as an expansion in powers of Newton's constant. 

The important point is that the set of correlation functions $\{{\cal G}_n\}$ can be shown to be correlators in a (non-unitary) conformal field theory: more specifically, each ${\cal G}_n$ is the expectation value of a product of $n$ energy-momentum tensors in the CFT (and, if there are matter fields, there are also the corresponding matter currents). In other words, the full set of correlation functions $\{{\cal G}_n\}$ determines the partition function of such a CFT. 

Thus a single quantum gravity wave-functional $\Psi[g]$ corresponds to the partition function of a non-unitary CFT, hence Chakraborty et al.'s notation $Z[g]$. Since the space of solutions of the WDW equation is the space of wave-functionals, which is the space of CFTs, this motivates us to call such a CFT a `model'.\footnote{I am here identifying a CFT with its partition function, which includes sources for operators. The reason is that all the correlation functions of these operators can be recovered from the partition function. This is analogous with the role of the partition function in Gibbsian statistical mechanics.}
The set of all such CFT partition functions is a theory space that makes up the state-space of quantum cosmology. 

\subsection{A normed vector space}

The above set of wave-functions $\Psi[g]$ was given in a perturbative expansion, around the Euclidean vacuum. This space of solutions comes with a natural norm. This norm is obtained by integrating $|Z[g]|^2$ over the spatial metrics, and dividing by the volume of its symmetry group, namely the diffeomorphism and Weyl groups.\footnote{Technically, this is done using the Faddeev-Popov procedure: see Chakraborty et al. (2023:~pp.~9--12).}
The explicit expression for this norm can be found in Chakraborty et al.~(2023:~pp.~4, 10). It is important to note that it is valid beyond the limit $\k\rightarrow0$. In particular, its $\k\rightarrow0$ limit reproduces the earlier norm written by Higuchi for quantum field theory in de Sitter space, in terms of de Sitter-invariant states. However, the Chakraborty et al.~(2023) norm goes beyond this limit.\footnote{There are some interesting subtleties about this limit that I cannot address here, and which signal significant differences between quantum gravity and quantum field theory. For a discussion, see Chakraborty et al.~(2023:~p.~21).}

The main upshot of this discussion is that it gives a vivid illustration, for a cosmologically relevant treatment of quantum cosmology, of the two main aspects of the geometric view. Namely, the partition function of a CFT is ``merely'' a wave-function, i.e.~one element of the state-space of de Sitter quantum gravity. Thus what we normally think of as being a theory, is here rightly thought of as a model. Second, there is further structure on the set of CFTs, i.e.~on the set of models: namely, it is a normed vector space. While this latter propery is part of the definition of a state-space, it is {\it not} in general to be expected of the set of CFTs. 
It is the special set of (non-unitary) CFTs that make up the theory of quantum cosmology, that can be given this vector space and normed structure.\footnote{It is only `normed' structure and not an inner product, because these papers do not work out the inner products between different vectors. Hence my calling it a state-space rather than a Hilbert space. For a general discussion of probability measures in the context of classical cosmology, see Curiel (2015).}

\end{appendices}

\section*{References}
\addcontentsline{toc}{section}{References}

\small

\ \\Atiyah, M.~F.~and Hitchin, N.~(1988). {\it The Geometry and Dynamics of Magnetic Monopoles}. Princeton: Princeton University Press.

\ \\Belavin, A.~A.~and Knizhnik, V.~G.~(1986). `Algebraic Geometry and the Geometry of Quantum Strings'. {\it Physics Letters} B, 168 (3), pp.~201--206.

\ \\Belot, G.~(2018). `Fifty Million Elvis Fans Can't be Wrong'. {\it Nous}, 52 (4), pp.~946--981.

\ \\Besse, A.~(1980). {\it Einstein Manifolds}. Berlin: Springer.

\ \\Chakraborty, T., Chakravarty, J., Godet, V., Paul, P.~and Raju, S.~(2023). `Holography of information in de Sitter space'. {\it Journal of High-Energy Physics}, 12, 120, pp.~1--44.

\ \\Chakraborty, T., Chakravarty, J., Godet, V., Paul, P.~and Raju, S.~(2024). `The Hilbert Space of de Sitter Quantum Gravity'. {\it Journal of High-Energy Physics}, 1 (132), pp.~1--44.

\ \\Curiel, E.~(2014). `Classical Mechanics Is Lagrangian; It Is Not Hamiltonian'. {\it The British Journal for the Philosophy of Science}, 65, pp.~269--321.

\ \\Curiel, E.~(2015). `Measure, topology and probabilistic reasoning in cosmology'. arXiv:1509.01878.

\ \\Cinti, E.~and S.~De Haro~(2026). `The Classical Limits of T-duality'. In preparation.

\ \\Dawid, R.~(2013). {\it String Theory and the Scientific Method}. Cambridge: Cambridge University Press.

\ \\De Boer, J.~and S.~De Haro~(2004). `The off-shell m5-brane and non-perturbative gauge theory'. {\it Nuclear Physics} B, 696, pp.~174--204.

\ \\De Haro, S.~(2017a). `Spacetime and Physical Equivalence'. In: {\it Beyond Spacetime. The Foundations of Quantum Gravity}, Huggett, N., Matsubara, K.~and W\"uthrich, C.~(Eds.), pp.~257--283. Cambridge: Cambridge University Press, 2020. 

\ \\De Haro, S.~(2017b). `Dualities and emergent gravity: Gauge/gravity duality'. {\it Studies in History and Philosophy of Modern Physics}, 59, pp.~109--125.

\ \\De Haro, S.~(2019). `Towards a Theory of Emergence for the Physical Sciences'. {\it European Journal for Philosophy of Science}, 9 (38), pp.~1--52.

\ \\De Haro, S. (2019a). `The heuristic function of duality'. {\it Synthese} , 196 (12), pp.~5169--5203. 

\ \\De Haro, S.~(2021). `Theoretical Equivalence and Duality'. {\it Synthese}, 198:~pp.~5139--5177.

\ \\De Haro, S.~and Butterfield, J.~N.~(2018). `A Schema for Duality, Illustrated by Bosonization'. In: {\it Foundations of Mathematics and Physics one Century after Hilbert}. Kouneiher, J.~(Ed.), pp.~305--376. Cham: Springer.

\ \\De Haro, S.~and Butterfield, J.~N.~(2025). {\it The Philosophy and Physics of Duality}. Oxford: Oxford University Press.

\ \\De Haro, S., Teh, N., Butterfield, J.N.~(2017). `Comparing dualities and gauge symmetries'. {\it Studies in History and Philosophy of Modern Physics}, 59, pp.~68--80.

\ \\Dewar, N.~(2019). `Sophistication about symmetries'. {\it The British Journal for the Philosophy of Science}, 70 (2), pp.~485--521.

\ \\Eisenbund, D.~and Harris, H.~(2000). {\it The Geometry of Schemes}. New York: Springer.

\ \\Eskens, F.~D.~J.~(2025). `On effective dualities'. {\it Synthese}, 206, 155, pp.~1--36.

\ \\Farkas, H.~M.~and Kra, I.~(1980). {\it Riemann Surfaces}. New York: Springer.

\ \\Feintzeig, B.~H.~(2020). `The Classical Limit as an Approximation'. {\it Philosophy of Science}, 87, pp.~612--639.

\ \\Feintzeig, B.~H.~(2022). {\it The Classical-quantum correspondence}. Cambridge: Cambridge University Press.

\ \\Fletcher, S.~C.~(2016). `Similarity, Topology, and Physical Significance in Relativity Theory'. {\it The British Journal for the Philosophy of Science}, 67, pp.~365--389.

\ \\Frigg, R.~(2023). {\it Models and theories}. London: Routledge.

\ \\Gaiotto, D.~and Witten, E.~(2022). `Gauge Theory and the Analytic Form of the Geometric Langlands Program'. {\it Annales Henri Poincar\'e}, pp.~1--115. Cham: Springer.

\ \\Geroch, R.~(1969). `Limits of Spacetimes'. {\it Communications in Mathematical Physics}, 13, pp.~180--193.

\ \\Giveon, A.~and Kutasov, D.~(1999). `Brane dynamics and gauge theory'. {\it Reviews of Modern Physics}, 71 (4), pp.~983--1084.

\ \\Glymour, C.~(2013). `Theoretical Equivalence and the Semantic View of Theories'. {\it Philosophy of Science}, 80, pp.~286--297.

\ \\Gomes, H.~(2025). `Gauge Theory Without Principal Fibre Bundles'. {\it Philosophy of Science},  92 (3), pp.~511--527.

\ \\Gukov, S.~and Witten, E.~(2006). `Gauge Theory, Ramification, and the Geometric Langlands Program'. {\it Current Developments in Mathematics}, 1, pp.~35--180.

\ \\Halvorson, H.~(2012). `What Scientific Theories Could Not Be'. {\it Philosophy of Science}, 79, pp.~183--206.

\ \\Halvorson, H.~(2019). {\it The Logic in Philosophy of Science}. Cambridge: Cambridge University Press.

\ \\Halvorson, H.~and Tsementzis, D.~(2017). `Categories of Scientific Theories'. In: Landry, E.~(Ed.), {\it Categories for the Working Philosopher}, pp.~402--429. Oxford: Oxford University Press.

\ \\Hamilton, M.~J.~D.~(2017). {\it Mathematical Gauge Theory}. Cham: Springer.

\ \\Huggett, N.~(2017), `Target space $\neq$ space'. {\it Studies in History and Philosophy of Modern Physics}, 59, pp.~81--88.

\ \\Huggett, N.~and W\"uthrich, C.~(2025). {\it Out of Nowhere}. Oxford University Press.

\ \\Isham, C.~J.~(1999). {\it Modern Differential Geometry for Physicists}. Singapore: World Scientific. Second Edition, 2001.

\ \\Jacobs, C.~(2023a). `The Metaphysics of Fibre Bundles'. {\it Studies in History and Philosophy of Science}, 97, pp.~34--43.

\ \\Jacobs, C.~(2023b). `The Nature of a Constant of Nature: The Case of $G$'. {\it Philosophy of Science}, 90, pp.~797--816.

\ \\Kapustin, A.~(2008). `Lectures on Electric-Magnetic Duality and the Geometric Langlands Program'. Lecture Notes Aarhus University, https://ctqm.au.dk/events/2007/August.

\ \\Kapustin, A.~and Witten, E.~(2007). `Electric-Magnetic Duality and the Geometric Langlands Program'. {\it Communications in Number Theory and Physics} 1, pp.~1--236.

\ \\Kodaira, K.~(1986). {\it Complex Manifolds and Deformation of Complex Structures}. Berlin: Springer, 2005 reprint.

\ \\Landsman, K.~(2017). {\it Foundations of 	Quantum Theory.} Cham: Springer.

\ \\Lehmkuhl, D.~(2017). `Introduction: Towards a Theory of Spacetime Theories'. In: {\it Towards a Theory of Spacetime Theories}, Lehmkuhl, D., Schiemann, G.~and Scholz, E.~(Eds.). New York: Springer.

\ \\Lutz, S.~(2017). `What Was the Syntax-Semantics Debate in the Philosophy of Science About?' {\it Philosophy and Phenomenological Research}, XCV (2), pp.~319--352.

\ \\Manton, N.~S.~(1982). `A Remark on the Scattering of BPS Monopoles. {\it Physics Letters} B, 110 B (1), pp.~54--56.

\ \\Manton, N.~and Sutcliffe, P.~(2004). {\it Topological Solitons}. Cambridge: Cambridge University Press.

\ \\Martens, N.~C.~and Read, J.~(2021). `Sophistry about symmetries?' {\it Synthese}, 199 (1), pp.~315--344.

\ \\Nakahara, M.~(2003). {\it Geometry, Topology and Physics}. Bristol: Institute of Physics.

\ \\Nelson, P.~(1987). `Lectures on strings and moduli space'. {\it Physics Reports}, 149 (6), pp.~337--375.

\ \\North, J.~(2021). {\it Physics, Structure, and Reality}. Oxford: Oxford University Press. 

\ \\Read, J.~(2016). `The Interpretation of String-Theoretic Dualities'. {\it Foundations of Physics}, 46, pp.~209--235.

\ \\Read, J.~and M\o ller-Nielsen, T.~(2020). `Motivating Dualities'. {\it Synthese}, 197, pp.~263--291.

\ \\Rickles, D.~(2011). `A Philosopher looks at String Dualities'. {\it Studies in History and Philosophy of Modern Physics}, 42 pp.~54--67.

\ \\Rickles, D.~(2013). `AdS/CFT duality and the emergence of spacetime'. {\it Studies in History and Philosophy of Modern Physics}, 44, pp.~312--320.

\ \\Rickles, D.~(2017). `Dual theories: `same but different' or different but same'?' {\it Studies in History and Philosophy of Modern Physics}, 59, 62-67.

\ \\Riemann, B.~(1857). `Theorie der Abel'schen Functionen'. {\it Journal f\"ur die reine und angewandte Mathematik}, 54, pp.~101--155. Transcribed by D.~R.~Wilkins, 2000.

\ \\Schlichenmaier, M.~(2007). {\it An Introduction to Riemann Surfaces, Algebraic Curves and Moduli Spaces}. Berlin: Springer.

\ \\Seiberg, N.~(1995), `Electric-magnetic duality in supersymmetric non-Abelian gauge theories'. {\it Nuclear Physics} B, pp.~129--146.

\ \\Seiberg, N.~and Shao, S.-H.~(2023). `Majorana chain and Ising model -- (non-invertible) translations, anomalies, and emanant symmetries'. {\it SciPost Physics}, 16 (3), 064.

\ \\Seiberg, N.~and Witten, E.~(1994a). `Electric-Magnetic Duality, Monopole Condensation, and Confinement in ${\cal N}=2$ Supersymmetric Yang-Mills Theory'. {\it Nuclear Physics} B, 426, pp.~19--52.

\ \\Seiberg, N.~and Witten, E.~(1994b). `Monopoles, Duality and Chiral Symmetry Breaking in ${\cal N}=2$ Supersymmetric QCD'. {\it Nuclear Physics} B, 431, pp.~484--550.

\ \\Shao, S.~H.~(2023). `What's done cannot be undone: TASI lectures on non-invertible symmetries'. ArXiv preprint arXiv:2308.00747.

\ \\Sharpe, E.~(2003). `Lectures on D-branes and Sheaves'. ArXiv: hep-th/0307245.

\ \\Steeger, J.~and Feintzeig, B.~H.~(2021). `Is the classical limit ``singular''?' {\it Studies in History and Philosophy of Science}, 88, pp.~263--279.

\ \\Vafa, C.~(1998). `Geometric Physics'. {\it Documenta Mathematica}, Extra Volume International Congress of Mathematicians, Volume I, pp.~537--556.

\ \\van Fraassen, B.~C.~(2014). `One or Two Gentle Remarks about Hans Halvorson's Critique of the Semantic View'. {\it Philosophy of Science}, 81, pp.~276--283.

\ \\Vergouwen, S.~and De Haro, S.~(2025). `Supersymmetry in the Seiberg-Witten Theory: A Window into Quantum Field Theory'. {\it Synthese}, 205 (60), pp.~1--24.

\ \\Vistarini, T.~(2019). {\it The Emergence of Spacetime in String Theory}. New York: Taylor and Francis.

\ \\Wallace, D.~(2022). `Stating Structural Realism: Mathematics-First Approaches to Physics and Metaphysics'. {\it Philosophical Perspectives}, 36, pp.~345--378.

\ \\Wallace, D.~(2024). `Learning to Represent: Mathematics-first accounts of representation and their relation to natural language'. https://philsci-archive.pitt.edu/23224.

\ \\Weatherall, J.~O.~(2019). `Theoretical equivalence in physics'. {\it Philosophy Compass}, 14. Part 1: https://doi.org/10.1111/phc3.12591. Part 2: https://doi.org/10.1111/phc3.12591.

\ \\Weatherall, J. O. (2020). `Equivalence and duality in electromagnetism'. {\it Philosophy of Science}, 87 (5),
pp.~1172--1183.

\ \\Witten, E.~(1997). `Solutions of four-dimensional field theories via M-theory'. {\it Nuclear Physics} B, 500, pp.~3--42.

\end{document}